\PassOptionsToPackage{prologue,dvipsnames,table}{xcolor}

\documentclass[sigconf, nonacm]{acmart}

\newcommand\vldbdoi{XX.XX/XXX.XX}
\newcommand\vldbpages{XXX-XXX}
\newcommand\vldbvolume{18}
\newcommand\vldbissue{1}
\newcommand\vldbyear{2025}

\newcommand\vldbtitle{\shorttitle} 
\newcommand\vldbavailabilityurl{https://github.com/WASSER2545/PBench}
\newcommand\vldbpagestyle{plain} 
\newcommand{\oursys}{\sys}
\newcommand{\sys}{\textsc{PBench}\xspace}
\newcommand{\snowset}{{\textit{Snowset}}\xspace}
\newcommand{\redset}{{\textit{Redset}}\xspace}

\usepackage{graphicx}
\usepackage{balance}  
\usepackage{tikz}
\usepackage{subcaption}
\usepackage{xcolor}
\usepackage{cleveref}
\usepackage[ruled,linesnumbered,noend]{algorithm2e}
\usepackage{adjustbox}
\usepackage{array}
\newcolumntype{C}[1]{>{\centering}p{#1}}
\usepackage{booktabs}
\usepackage{multirow}
\usepackage{enumitem}
\usepackage{bbold}
\usepackage{tcolorbox}
\usetikzlibrary{shapes, positioning, arrows.meta}
\usepackage{soul}
\newcommand{\stitle}[1]{\vspace{1mm}\noindent{\bf #1}}
\newcommand{\etitle}[1]{\vspace{1mm}\noindent{\underline{\em #1}}}

\usepackage[noend]{algpseudocode}
\usepackage{float}
\usepackage{lipsum}
\usepackage{makecell} 


\usepackage{tikz}
\usepackage{framed}

\setcopyright{none}
\renewcommand\footnotetextcopyrightpermission[1]{} 
\begin{document}
\title{\oursys: Workload Synthesizer with Real Statistics for Cloud Analytics Benchmarking}

\settopmatter{authorsperrow=4}
\author{Yan Zhou\texorpdfstring{$^*$}{} }
\affiliation{%
  \institution{Renmin University, China}
}
\orcid{0009-0001-3412-1922}
\email{zhouyan2018@ruc.edu.cn}

\author{Chunwei Liu\texorpdfstring{$^*$}{} }
\orcid{0000-0003-1481-2678}
\affiliation{%
  \institution{MIT CSAIL}
}
\email{chunwei@csail.mit.edu}

\author{Bhuvan Urgaonkar}
\affiliation{%
  \institution{Penn State and AWS}
}
\email{urgaonkb@amazon.com}

\author{Zhengle Wang}
\affiliation{%
  \institution{Renmin University, China}
}
\orcid{0009-0003-0221-6761}
\email{wangzhengle@cau.edu.cn}

\author{Magnus Mueller}
\affiliation{%
  \institution{Amazon Web Services}
}
\email{magnusmu@amazon.com}

\author{Chao Zhang}
\affiliation{%
  \institution{Renmin University, China}
}
\email{cycchao@ruc.edu.cn}

\author{Songyue Zhang}
\affiliation{%
  \institution{Renmin University, China}}
\email{zhangsongyue@ruc.edu.cn}

\author{Pascal Pfeil}
\affiliation{%
  \institution{Amazon Web Services}
}
\email{pfeip@amazon.de}

\author{Dominik Horn}
\affiliation{%
  \institution{Amazon Web Services}
}
\email{domhorn@amazon.de}

\author{Zhengchun Liu}
\affiliation{%
  \institution{Amazon Web Services}
}
\email{zcl@amazon.com}

\author{Davide Pagano}
\affiliation{%
  \institution{Amazon Web Services}
}
\email{dpagano@amazon.com}
    
\author{Tim Kraska}
\affiliation{%
  \institution{MIT and AWS}
}
\email{kraska@mit.edu}

\author{Samuel Madden}
\affiliation{%
  \institution{MIT CSAIL}
}
\email{madden@csail.mit.edu}

\author{Ju Fan}
\affiliation{%
  \institution{Renmin University, China}
}
\email{fanj@ruc.edu.cn}


\begin{abstract}
Cloud service providers commonly use standard benchmarks like TPC-H and TPC-DS to evaluate and optimize cloud data analytics systems. However, these benchmarks rely on fixed query patterns and fail to capture the real execution statistics of production cloud workloads.
Although some cloud database vendors have recently released real workload traces, these traces alone do not qualify as benchmarks, as they typically lack essential components like the original SQL queries and their underlying databases.
To overcome this limitation, this paper introduces a new problem of \emph{workload synthesis with real statistics}, which aims to generate \emph{synthetic workloads} that closely approximate real execution statistics, including key performance metrics and operator distributions, in real cloud workloads.
To address this problem, we propose \sys, a novel workload synthesizer that constructs synthetic workloads by judiciously selecting and combining workload components (i.e., queries and databases) from existing benchmarks.
This paper studies the key challenges in \sys. First, we address the challenge of balancing performance metrics and operator distributions by introducing a multi-objective optimization-based component selection method. Second, to capture the temporal dynamics of real workloads, we design a timestamp assignment method that progressively refines workload timestamps. Third, to handle the disparity between the original workload and the candidate workload, we propose a component augmentation approach that leverages large language models (LLMs) to generate additional workload components while maintaining statistical fidelity.
%
%
%
%
We evaluate \sys on real cloud workload traces, demonstrating that it reduces approximation error by up to $6{\small \times}$ compared to state-of-the-art methods.
\end{abstract}

\maketitle

\pagestyle{\vldbpagestyle}
\begingroup\small\noindent\raggedright\textbf{PVLDB Reference Format:}\\
\vldbtitle. PVLDB, \vldbvolume(\vldbissue): \vldbpages, \vldbyear.\\
\href{https://doi.org/\vldbdoi}{doi:\vldbdoi}
\endgroup
\begingroup
\renewcommand\thefootnote{}
\footnote{\noindent \textsuperscript{*} indicates equal first author contribution.\\
This work is licensed under the Creative Commons BY-NC-ND 4.0 International License. Visit \url{https://creativecommons.org/licenses/by-nc-nd/4.0/} to view a copy of this license. For any use beyond those covered by this license, obtain permission by emailing \href{mailto:info@vldb.org}{info@vldb.org}. Copyright is held by the owner/author(s). Publication rights licensed to the VLDB Endowment. \\
\raggedright Proceedings of the VLDB Endowment, Vol. \vldbvolume, No. \vldbissue\ %
ISSN 2150-8097. \\
\href{https://doi.org/\vldbdoi}{doi:\vldbdoi} \\
}\addtocounter{footnote}{-1}\endgroup

\ifdefempty{\vldbavailabilityurl}{}{
\vspace{.3cm}
\begingroup\small\noindent\raggedright\textbf{PVLDB Artifact Availability:}\\
The source code, data, and/or other artifacts have been made available at \url{\vldbavailabilityurl}.
\endgroup
}

\section{Introduction}
\label{sec:intro}

The growing significance of cloud-based database systems has created a demand for new {benchmarking} approaches.
Existing benchmarks, notably those from TPC~\cite{tpc-h-vldb,tpc-ds, boncz2013tpc}, are not well-suited for cloud workloads due to two key limitations. 
First, these benchmarks are not designed to accommodate the varying degrees of concurrency commonly observed in cloud databases, making them ineffective in accurately capturing the \emph{performance characteristics} of modern cloud database systems.
Second, recent studies~\cite{CAB, redset} have shown that the distribution of query operators (e.g., the frequency of joins and aggregates) in modern cloud workloads differs significantly from those in traditional benchmarks. Given that different operator distributions correspond to various business logic patterns and can lead to varying cost-optimal query optimization strategies in the cloud~\cite{dong2024cloud, leis2021towards, zhang2023cost}, it is crucial for benchmark workloads to accurately reflect the real-world \emph{operator distributions}.

\begin{figure}[t!]
    \centering
    \includegraphics[width=0.99\linewidth]{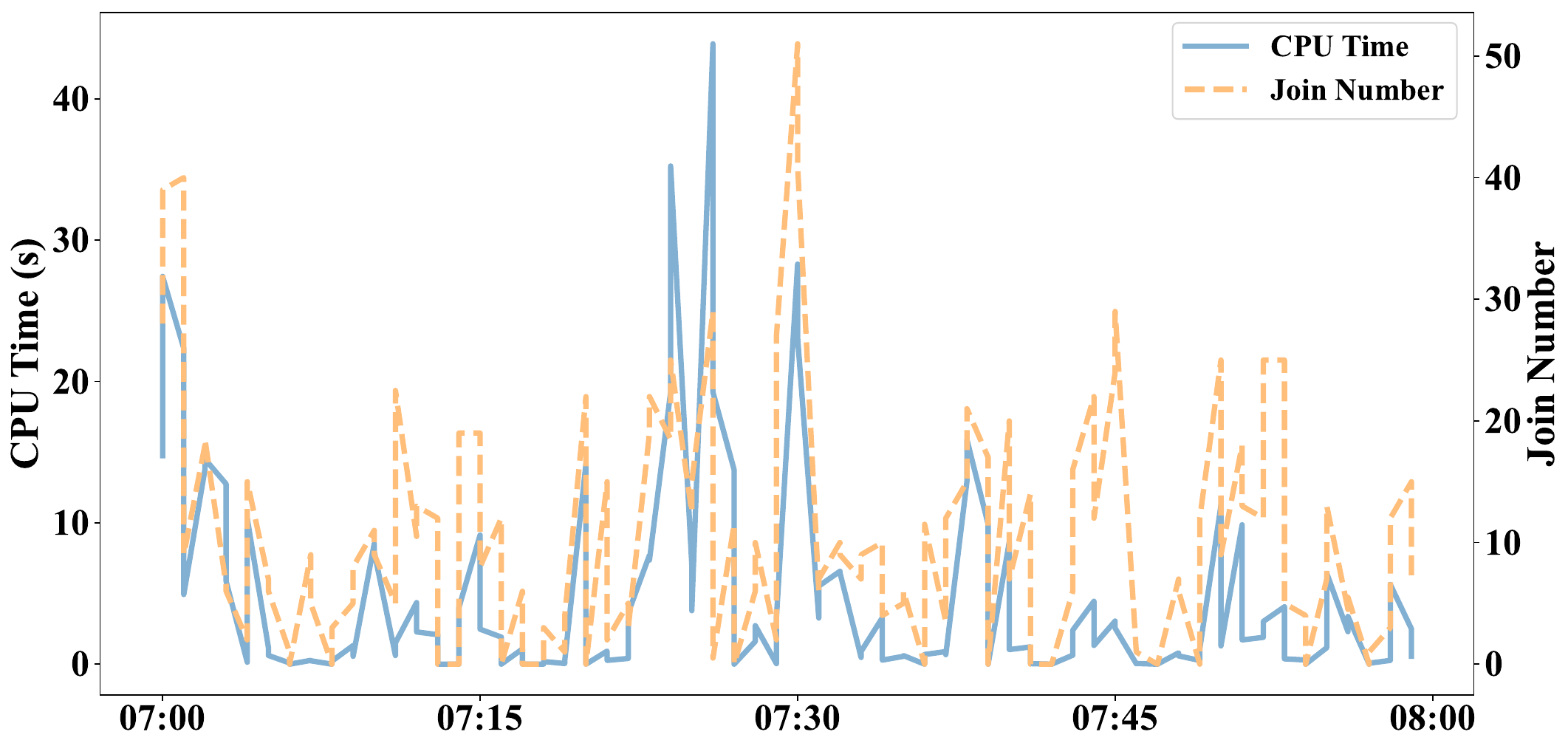}
    \vspace{-1em}
    \caption{An Example Workload Trace from \redset.}
    \vspace{-1em}
    \label{fig:redset-trace}
\end{figure}

To address these limitations, several cloud database vendors, including Snowflake~\cite{snowflake} and Amazon Redshift~\cite{armenatzoglou2022amazon}, have released \emph{customer workload traces}~\cite{redset,snowset}, which are anonymized logs capturing \emph{real execution statistics} from cloud workloads. These traces include key performance metrics (e.g., CPU Time) and workload characteristics (e.g., operator distributions), offering valuable insights into real-world cloud database behavior. 
Figure~\ref{fig:redset-trace} illustrates an example workload trace from \redset~\cite{redset}, which provides real statistics, including a performance metric, CPU Time, and an operator distribution feature, Join Number.

Leveraging these traces presents an opportunity to develop benchmarks that better reflect real-world cloud database workloads. However, these traces alone do not constitute benchmarks, as they typically lack essential components like the original SQL queries and their underlying databases, making them unsuitable for directly evaluating cloud database system performance.

\stitle{Workload Synthesis with Real Statistics.}
To address this issue, this paper introduces a new problem of \emph{workload synthesis with real statistics}, which aims to generate \emph{synthetic workloads} that closely approximate the {performance metrics} (e.g., CPU Time) and {operator distributions} (e.g., the frequency of joins, scans, and aggregations) observed in real cloud workload traces.
Given that real workload traces lack the original SQL queries and their underlying databases, this work focuses on constructing synthetic workloads by combining \emph{workload components}, i.e., SQL queries and their corresponding databases, from \emph{existing benchmarks}, such as TPC-H and TPC-DS.
The goal is to construct a synthetic workload that minimizes the divergence in execution statistics, including performance metrics and operator distributions, from the real workload trace.
\begin{example}
	Consider an example workload trace from Redset~\cite{redset} in Figure~\ref{fig:redset-trace}. Since the original SQL queries and their corresponding databases are not provided in the trace, we construct a synthetic workload by selecting workload components, such as 30\% of queries from TPC-H and 70\% from TPC-DS. We then carefully assign timestamps to these selected queries to approximate the temporal characteristics of the real workload. After executing the synthetic workload, we obtain execution statistics (CPU Time and Join Number), and our goal is to minimize the divergence between these synthetic statistics and the real statistics in Figure~\ref{fig:redset-trace}.
\end{example}

\stitle{Key Challenges.}
Effectively addressing the problem of \emph{workload synthesis with real statistics} presents three key technical challenges.

\etitle{(C1) Balancing approximation objectives}.
Selecting an appropriate set of workload components from existing benchmarks is challenging due to the need to simultaneously approximate both performance metrics and operator distributions.
These two objectives are inherently intertwined, i.e., adjusting one (e.g., increasing the join number) often impacts the other (e.g., leading to higher CPU time). This interdependence complicates the workload component selection process, making it difficult to construct a synthetic workload that accurately captures the statistics of real workload traces.

\etitle{(C2) Preserving temporal dynamics}.
Since cloud workloads exhibit varying levels of concurrency, short-term fluctuations (e.g., within 30 seconds), and periodic query patterns~\cite{li2022cloud}, naively distributing queries over time may fail to capture these temporal dynamics, leading to discrepancies between the synthetic and real workloads.
Thus, the second challenge lies in effectively assigning timestamps to the selected workload components to preserve the temporal dynamics in real workload traces. 


\etitle{(C3) Augmenting workload components.}
Workload components from existing benchmarks may be insufficient to construct a synthetic workload that accurately reflects real traces due to the disparity between the original workload and the candidate workload. These disparity are often caused by different query patterns and underlying databases, and it is challenging to augment additional workload components of high quality due to the huge search space.

%

\stitle{Our \sys Approach.}
In this paper, we propose \sys, a novel approach for workload synthesis that effectively approximates real statistics from cloud workload traces. \sys takes a real workload trace (e.g., from Redset~\cite{redset} or Snowset~\cite{snowset}) as input and constructs a synthetic workload by strategically selecting and combining workload components from existing benchmarks. 

Technically, to address challenge (C1), we propose a \textbf{component selection} approach that effectively balances the approximation of both performance metrics and operator distributions. To achieve this, we formulate component selection as a multi-objective optimization problem using integer linear programming (ILP) and employ efficient algorithms to solve it.
To address challenge (C2), we propose a \textbf{timestamp assignment} method that incrementally refines the temporal distribution of the selected workload components. Specifically, we employ a simulated annealing-based approach that progressively adjusts workload timestamps, transitioning from a coarse-grained approximation to a fine-grained alignment with the temporal dynamics of real workload traces.
To address challenge (C3), we propose a \textbf{component augmentation} approach that leverages large language models (LLMs) to generate additional workload components. To this end, we design effective prompting strategies and a trial-and-error generation mechanism for the LLM to produce new workload components that significantly reduce the approximation error.

\stitle{Differences from Existing Methods}. 
Several existing studies have explored generating synthetic workloads based on standard benchmarks~\cite{Stitcher,CAB}. However, they fall short in accurately approximating complex real cloud workload traces.
First, CAB~\cite{CAB} employs heuristic strategies, such as randomly selecting queries from a predefined query pool, and lacks a principled method to ensure that the generated synthetic workload closely approximates the real workload.
Second, while Stitcher~\cite{Stitcher} employs Bayesian optimization~\cite{frazier2018tutorial}, it overlooks operator distributions. Given that  the inherent interaction between performance metrics and operator distributions, as mentioned previously, this limitation hinders Stitcher’s ability to generate balanced workloads. Moreover, Stitcher selects all queries from a single benchmark without filtering out irrelevant or redundant queries, thus lacking a \emph{fine-grained} mechanism to select queries relevant to real workloads.

\stitle{Contributions.} Our contributions are summarized as follows.

\vspace{1mm} \noindent
(1) We introduce a novel problem of workload synthesis with real statistics, which is formally defined in Section~\ref{sec:preliminaries}.

\vspace{1mm} \noindent
(2) We propose \sys, an effective approach to workload synthesis with real statistics (Section~\ref{sec:framework}). We develop effective techniques in \sys for component selection and timestamp assignment (Section~\ref{sec:two_phase}). We also leverage an LLM-enhanced method for component augmentation (Section~\ref{sec:LLM_gen}).

\vspace{1mm} \noindent
(3) We developed and open-sourced \sys to support cloud analytics benchmarking. To evaluate its effectiveness, we conducted extensive experiments on two widely-recognized cloud workload traces, \snowset~\cite{snowset} and \redset~\cite{redset} (Section~\ref{sec:eval}). Experimental results show that \sys achieves up to a $6{\small \times}$ reduction in approximation error compared to state-of-the-art methods.

\begin{table}[!t] 
    \small
    \caption{Notations }
    \vspace{-1.2em}
    \label{tab:notations}
    \setlength\tabcolsep{3pt}
  \resizebox{\linewidth}{!}{
        \begin{tabular}{c|c||c|l} \toprule
        \textbf{Notation} & \textbf{Description} & \textbf{Notation} & \textbf{Description} \\ \hline
\(W\) & Original Workload &  $q_j'$ & Query of $C_j$ \\ \hline
\(\widetilde{W}\) & Synthesized Workload  & $D_j'$& Database of $C_j$\\ \hline
$F$ & Performance Feature & $F_j'$&  $F$ of $C_j$\\ \hline
\(M\) & Performance Metrics  & $m_{hj}'$&  $h$-th $M$ of $C_j$ \\  \hline
\(O\) & Operator Distributions& $o_{uj}'$& $u$-th $O$ of $C_j$ \\ \hline
$q_j$& $j$-th Customer Query& $T_j'$& Duration of $C_j$  \\ \hline
$t_j$ & Timestamp of $q_j$ & \(\widetilde{W_C}\)&  Workload by \(\mathbb{C}\)\\ \hline
$D$& Customer's Database& $x_j^{(i)}$ &  Number of $C_j$ in $\Gamma^{(i)}$ of \(\widetilde{W_C}\)\\ \hline
$F_j$& $F$ of $q_j$& $t_{jk}^{(i)}$&  $t$ of $k$-th $C_j$ in $\Gamma^{(i)}$ of \(\widetilde{W_C}\)\\ \hline
$\Gamma^{(i)}$ & \(i\)-th Time Window & $F_j^G$ & $j$-th Generation Goal \\ \hline
 $\gamma^{(ij)}$ &  $\Gamma^{(i)}$ 's $j$-th Time Interval & $\mathbb{E_{P}}_i$ & $F_j^G$ 's Positive Example \\ \hline
$F^{(i)}$ & $W$'s $F$ in $\Gamma^{(i)}$& $\mathbb{E_{N}}_j$ & $F_j^G$ 's Negative Examples \\\hline
$\widetilde{F^{(i)}}$& \(\widetilde{W}\)'s $F$ in $\Gamma^{(i)}$ & $C_{ij}^P$& $F_i^G$ 's $j$-th Positive Examples\\
\hline
 $F^{(ij)}$& $W$'s $F$ in $\gamma^{(ij)}$& $C_{ij}^N$&$F_i^G$ 's $j$-th Negative Example\\ \hline
 $\widetilde{F^{(ij)}}$& \(\widetilde{W}\)'s $F$ in $\gamma^{(ij)}$ & $\widetilde{C}$ &Synthesized Component \\ \hline
 $C_j$ & $j$-th Workload Component  & $F_{\widetilde{C}}$& $F$ of $\widetilde{C}$\\ \bottomrule
        \end{tabular}
    }
  \vspace{-1.2em}
\end{table} 

\section{Preliminaries}
\label{sec:preliminaries}
This section first some presents background information on real cloud workload traces~\cite{redset,snowset}, which are released by cloud database vendors Snowflake~\cite{snowflake} and Amazon Redshift~\cite{armenatzoglou2022amazon}, in Section~\ref{subsec:trace-background}. We then formally define the problem of workload synthesis with real statistics in Section~\ref{subsec:problem}.

\subsection{Cloud Workload Traces}
\label{subsec:trace-background}
Due to privacy concerns, customer's SQL queries and their underlying databases are typically unavailable. Thus, cloud vendors can only release workload traces containing aggregated statistics, such as performance metrics (e.g., CPU Time and Scanned Bytes) and operator distributions (e.g., operator types and their counts). Due to this reason, generating synthetic workloads that approximate these \emph{real} statistics is valuable for benchmarking and optimizing cloud databases, which motivates us to study the problem of workload synthesis with real statistics. 
Below, we describe two widely-recognized cloud workload traces, \snowset~\cite{snowset} and \redset~\cite{redset}.

\stitle{\snowset \cite{snowset}} is a real cloud workload trace released by Snowflake~\cite{snowflake}, a widely used cloud data warehouse. It captures approximately 69 million queries executed across Snowflake clusters over two weeks. While \snowset does not include original SQL queries or data, it provides per-query runtime statistics, including query submission time, cluster and database association, CPU usage, scanned bytes, and execution times for various operator types.
A detailed analysis conducted in~\cite{CAB} highlights key differences between \snowset and traditional TPC benchmarks in terms of query types, operator distributions, and resource consumption patterns. Notably, \snowset\ exhibits a significant skew in CPU Time, where nearly half (45.6\%) is consumed by a small number of large queries. Moreover, in contrast to TPC-H~\cite{tpc-h-vldb} and TPC-DS~\cite{tpc-ds}, which allocate a larger proportion of execution time to join operations, \snowset demonstrates a more diverse operator distribution, better reflecting real-world cloud workloads.




\stitle{\redset~\cite{redset}} is a real cloud workload trace recently released by Amazon, capturing query metrics collected over three months from 200 randomly selected Amazon Redshift~\cite{armenatzoglou2022amazon} serverless and provisioned instances. 
A comparative study in~\cite{redset} analyze differences between Redshift user workloads and TPC benchmark suites at both the query and workload levels. Notably, \redset queries exhibit significantly longer execution times and an extremely long-tailed distribution compared to TPC-DS. Furthermore, \redset demonstrates a high degree of query runtime skew, with less than 0.1\% of queries consuming approximately 25\% of all computing resources, whereas TPC-DS queries follow a more uniform distribution.




\stitle{Implications of \snowset and \redset.}
\snowset\ and \redset\ share key similarities while also exhibiting differences. Both datasets provide real execution statistics of real workloads without exposing original SQL queries and customers' databases. In particular, the execution statistics contain performance metrics that measure resource consumption (such as CPU Time and scanned bytes) and operator distributions. However, they differ in the granularity of query operator information: \snowset records the execution time of each operator, whereas \redset reports only operator counts.


%
The above characteristics of real workload traces from \snowset and \redset motivates us to study a new problem of workload synthesis with real statistics. 
Unlike traditional cloud vendor tools, which require access to original SQL queries and customers' databases, \sys aims to generate synthetic workloads that closely approximate the performance metrics and operator distributions, without accessing to queries and databases.
Thus, this capability allows datasets like \snowset and \redset to be leveraged more effectively for research on cloud database evaluation and optimization.



\subsection{Problem Formalization}
\label{subsec:problem}

\begin{figure}
    \centering
    \includegraphics[width=\linewidth]{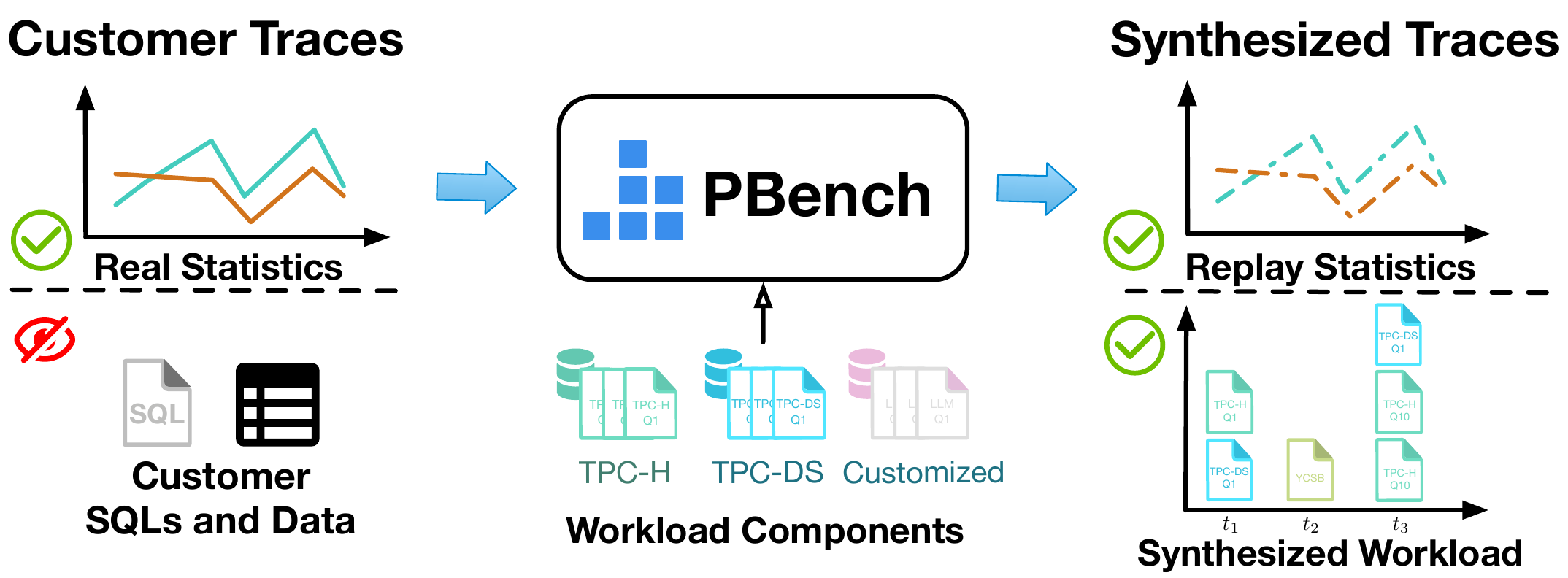}
        \vspace{-2em}
    \caption{Workload Synthesis with Real Statistics.
    }
    \vspace{-1em}
    \label{fig:overview}
\end{figure}



Figure~\ref{fig:overview} illustrates the problem of workload synthesis with real statistics. Given a real workload trace where original SQL queries and underlying databases are unavailable, the problem is to generate a synthetic workload that, when executed, produces statistics closely approximate the target workload’s performance metrics and operator distributions.
In particular, given that real workload traces lack the original SQL queries and their underlying databases, this work focuses on constructing synthetic workloads by combining \emph{workload components}, i.e., SQL queries and their corresponding databases, from \emph{existing benchmarks}, such as TPC-H and TPC-DS.
Then, the generated synthetic workload offers an effective solution for database benchmarking~\cite{deep2021diametrics,zhang2024hybench,zhang2024htap}, performance tuning~\cite{ottertune}, and bug detection~\cite{ba2024keep}, allowing database developers to perform realistic evaluations while preserving customer privacy and avoiding exposure of sensitive workloads or data.

This paper considers two settings to workload synthesis: \emph{query-level} and \emph{window-level}. The query-level setting synthesizes a workload by generating a corresponding query for each original query, preserving the number of queries and their arrival times. However, this method may introduce significant computational overhead and higher approximation errors when identifying suitable queries for synthesis. In contrast, the window-level setting aims to approximate the aggregated statistics of all customer queries within a given time window, offering greater flexibility and reducing overall errors compared to the query-level setting.
Although we investigate both settings, this paper primarily focuses on the window-level approach due to space constraints. Note that our proposed approach can be easily extended to the query-level setting.

Below, we formally define the studied problem. The notations used throughout this paper are summarized in Table~\ref{tab:notations}.
We first formalize real workload traces derived from \snowset or \redset.

\begin{definition}
\textbf{Cloud Workload $W$.} A workload $W$ is a time series of queries \{$q_1$,\dots, $q_n$\} to be run against a database $D$, where each timestamped query $q_i$ is denoted as $\{D,\{t_i,q_i\}\}$.
Note that while the queries and their corresponding databases are conceptually represented, they remain inaccessible in practice.
\end{definition}

\begin{definition}
\textbf{Performance Feature $F$.} 
We formalize the real statistics of a workload trace using the performance feature $F=<M,O>$, where $M$ represents a set of performance metrics, and $O$ denotes operator distributions, capturing the percentage distribution of different SQL operators. 
To facilitate window-level synthesis, we aggregate the performance features within the \textit{i}-th time window $\Gamma^{(i)}$, yielding the window-based performance feature representation $<\Gamma^{(i)},M^{(i)},O^{(i)}>$. Here, $M^{(i)}$ is the aggregated performance metrics vector, and $O^{(i)}$ is the aggregated operator distribution vector.
%
\end{definition}

Recall that this work focuses on constructing synthetic workloads by combining \emph{workload components} from \emph{existing benchmarks}.
Thus, we define the \emph{workload component} as follows.
\begin{definition}
\textbf{Workload Component $C$.} The synthetic workload is constructed over a set of workload components ${ C_1,C_2,\dots,C_n}$, where each component $C_j$ includes a query $q_j'$ and its corresponding populated database $D_j'$, both derived from existing benchmarks.
\end{definition}

Now, we are ready to define the problem of workload synthesis with real statistics as follows.
\begin{definition}
\textbf{Workload Synthesis with Real Statistics.} 
Given the performance feature $F$ of a target workload $W$, the problem is to synthesize a workload $\widetilde{W}$ such that each window-based performance feature $<\Gamma^{(i)},\widetilde{M^{(i)}},\widetilde{O^{(i)}}>$ closely approximates $<\Gamma^{(i)},M^{(i)},O^{(i)}>$, minimizing the approximation error, e.g., measured using Mean Absolute Error (MAE).
%
\end{definition}







\noindent \textbf{Remarks.} 
We assume that the source cloud cluster and the target cloud cluster for benchmarking are sufficiently similar in terms of node count, machine configurations, and database system, ensuring a high-fidelity evaluation environment. Notably, our problem formalization is highly extensible, supporting the approximation of various performance metrics (e.g., CPU usage, memory consumption, scanned bytes, and IOPS) and diverse operator types (e.g., selection, projection, join, and aggregation).

\section{An Overview of \sys}
\label{sec:framework}
This section presents an overview of \sys.
We first introduce the overall framework of \sys in Section~\ref{subsec:framework}.
Next, we describe the pre-processing stage, \textit{synthesizer data preparation}, which prepares the necessary inputs for workload synthesis, in Section~\ref{subsec:data-prep}.

\subsection{Overall Framework of \sys}
\label{subsec:framework}

\begin{figure}
    \centering
    \includegraphics[width=\linewidth]{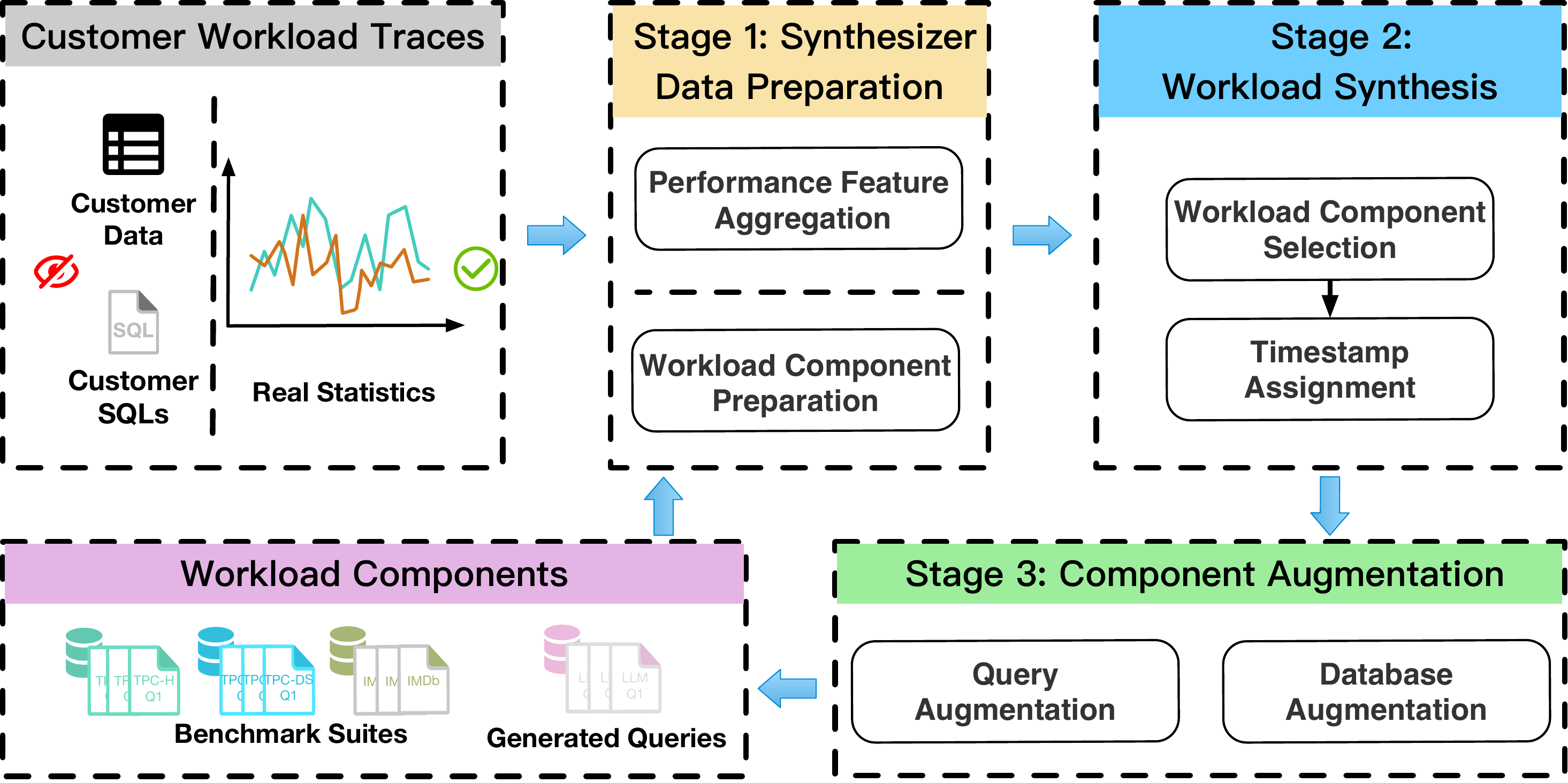}
    \caption{Overall Framework of \sys that consists of three stages: (1) \sys prepares workload components from standard benchmarks. (2) \sys employs a two-phase workload synthesis approach to generate the workload by selecting and combining components to approximate real workload statistics. (3) \sys augments workload components to further improve approximation effectiveness.
%
    	}
    \vspace{-1em}
    \label{fig:framework}
\end{figure}

Figure~\ref{fig:framework} illustrates the overall framework of \sys, which synthesizes a workload that closely approximates the performance metrics $M$ and operator distributions $O$ of a real workload $W$ trace through the following three stages:
(1) \textbf{Synthesizer Data Preparation}, where workload components are collected from standard benchmarks.
(2) \textbf{Two-Phase Workload Synthesis}, which first formulates the synthesis problem as an integer linear programming task and then assigns timestamps to preserve the temporal dynamics of the real workload $W$.
(3) \textbf{Workload Component Augmentation}, where an LLM-based module generates additional queries to further enhance approximation effectiveness.
Below, we provide a detailed explanation of each of the three stages in \sys.

%
%


\subsubsection{Synthesizer Data Preparation.} 
We select widely-used benchmarks for workload synthesis, including TPC-H, TPC-DS, and YCSB. TPC-H and TPC-DS serve as medium (22 query templates) and heavy (99 query templates) analytical workload components, respectively, while YCSB represents lightweight transactional workload components.
To prepare workload components, we consider two key factors: scale factor (\textit{SF}) and \textit{skewness}. The \textit{SF} parameter controls the database size, where a larger \textit{SF} leads to higher resource consumption for queries. The \textit{skewness} parameter controls dataset skewness, influencing query execution patterns. For instance, a higher skew results in high ratios between CPU Time and Scanned Bytes.
Once the workload components are prepared, we profile the queries to obtain their corresponding performance features. These features are aggregated at multiple levels to define the objectives for the workload synthesis tool, including: (1) Window-level aggregation to capture overall workload statistics, (2) Interval-level aggregation to reflect finer-grained execution patterns, and (3) Query-level features to model individual query characteristics.
For additional details on this stage, please refer to Section~\ref{subsec:data-prep}.

%
%


\subsubsection{Two-Phase Workload Synthesis.} 
Given a target workload trace $W$, we divide it into multiple time windows and aggregate performance metrics and operator distributions within each window. Thus, these aggregated statistics within each window (i.e., the aggregated performance features) can be regarded as the approximation objectives, e.g., total CPU Time and the count of join operators, as shown in Figure~\ref{fig:redset-trace}.
Then, we design a two-phase workload synthesis approach.
First, we formulate the workload synthesis problem as an optimization task that selects and combines existing workload components to approximate performance features of each time window $\Gamma^{(i)}$. 
To solve this optimization problem, we employ integer linear programming to determine the optimal combination of workload components.
Next, while the first phase generates a coarse-grained workload for each time window, it does not precisely distribute workload components over time. To achieve this, we introduce a timestamp assignment mechanism, which assigns a timestamp $t_{jk}^{(i)}$ for each instance of workload component $C_j$ within time window $\Gamma^{(i)}$. Since finding an optimal solution for this problem is computationally expensive, we design a Simulated Annealing (SA)-based algorithm to iteratively refine the timestamp assignment, improving the fidelity of the synthesized workload.
For additional details on this stage, please refer to Section~\ref{sec:two_phase}.

\subsubsection{Workload Component Augmentation.} 
Simply selecting workload components from existing benchmarks (like TPC-H and TPC-DS) is often insufficient to approximate the performance features of real workloads due to differences between the original workload and the candidate workload. To bridge this gap, we design an LLM-enhanced workload augmenter that automatically generates additional workload components to better approximate the real workload.
The high-level idea is first identifying underperforming time windows, then guiding the LLMs to generate additional components for these time windows, and finally enabling query refinement by injecting expert hints.
%
%
%
Selecting an appropriate database for the generated queries poses another challenge. Even though we prepare a set of benchmark databases in advance, discrepancies in data distribution between the generation target and the available databases may hinder the LLM’s ability to generate queries with the desired performance characteristics. To address this, we implement a heuristic-driven database selection mechanism that dynamically re-selects or regenerates databases when the LLM fails to produce a suitable query after multiple attempts.
%
For additional details on this stage, please refer to Section~\ref{sec:LLM_gen}.

\subsection{Synthesizer Data Preparation}
\label{subsec:data-prep}




The Synthesizer Data Preparation stage comprises two key processes: workload trace preparation and workload component preparation, which are discussed separately below.

\subsubsection{Workload Trace Preparation}
The statistics collected from the cloud may be query-level statistics (like \snowset\ or \redset) of one specific cluster and database or instance-level time series data which are denoted as $<q_i,F_i,t_i>$, where $q_i$ is inaccessible. When handling query-level statistics, it is required to set a generation granularity and aggregate query-level statistics into performance features $F^{(ij)}$. To achieve the synthesis goal that approximates the statistics at the time interval level, we need to aggregate them by assuming a uniform distribution. This assumption is necessary because datasets like \snowset\ and \redset\ do not provide the actual performance consumption of queries during their execution, making it inevitable to use assumptions like uniform distribution to derive each time interval's statistics. Regarding the selection of time window length, a larger time window results in an overall improvement in the time window by increasing the selection space of the first phase of workload synthesis. However, it increases the time consumption and decreases the accuracy of the second phase. To determine an appropriate time window length, we employ an empirical approach to balance generation accuracy and generation efficiency. After finalizing a time window, we aggregate the feature $F$ of all queries running within the time window $\Gamma^{(i)}$ to obtain the corresponding generation target $F^{(i)}$. The aggregation function may vary depending on the indicators or input datasets. For example, as for performance indicators like CPU Time and Scanned Bytes, we can sum up the resource consumption of all queries. For operator distribution, since \snowset\ provides information about operator runtime and \redset\ provides the number of operators for each query, we can aggregate and transform them into statistics using average or sum, thus we obtain the operator distribution of \snowset\ and operator numbers of \redset.   





\subsubsection{Workload Component Preparation}
To ensure a diverse set of queries and databases, we first generate benchmark databases with different sizes and data skewnesses. The database size is controlled by the \textit{scale factor} parameter and the data skewness can be generated using five scales of skewness utilizing tools like TPC-H skewed generation tool~\cite{skewtpch}~\cite{dsb}. The data size primarily impacts the Scanned Bytes; while the skewness will influence the joining fan-out distributions, thereby affecting the intermediate result size and the CPU Time of the queries. To compensate for the difference between the customer's invisible database and the pre-prepared benchmark databases in real-time, we propose a workload component augmentation method (See \autoref{sec:LLM_gen}).





To obtain the performance feature $F_j'$ of each workload component $C_j$, we profile these benchmark queries against their corresponding databases. Since different cache conditions will affect the duration of a query, we repeat running each query three times to obtain its features under different cache conditions and average the results. We observed that the query features from different benchmarks exhibit significant differences. For instance, TPC-DS queries typically contain more join and aggregation operators than TPC-H queries, and the former follows a star schema while the latter adheres to a third normal form. Therefore, it is essential to prepare a diverse set of workload components for effective query generation (See \autoref{sec:LLM_gen}).

\section{Two-Phase Workload Synthesis}
\label{sec:two_phase}





\subsection{Workload Component Selection Phase}



\subsubsection{Optimization Objective}
To synthesize a workload with real statistics, we combine multiple workload components within a time window, and each selected component can be repeated multiple times. Therefore, the objective is to select and combine multiple workload components to build $\widetilde{W_C}$ such that the total relative error of the performance feature is minimized. 




\begin{equation}
\label{equ:optimization_goal}
\begin{aligned}
\mathop{\arg\min}\limits_{x_j^{(i)}} \ \sum_{i=1}^n(\sum_{h=1}^{n_h} \left| \frac{\sum_{j=1}^{v} x_j^{(i)}*m_{hj}'-M_h^{(i)}}{M_h^{(i)}} \right|+ \\\sum_{u=1}^{n_u} \left| \frac{\sum_{j=1}^{v} x_j^{(i)}*o_{uj}'-O_u^{(i)}}{O_u^{(i)}} \right|)
\\
s.t. x_j^{(i)}\in \mathbb{Z^*},x_j^{(i)}<y,\sum_{j=1}^{v} x_j^{(i)}<z,\sum_{j=1}^{v} x_j^{(i)}*T_{j}'\leq l
\end{aligned}
\end{equation}

\noindent The optimization objective is denoted as the above equation, where $x_j^{(i)}$ is the repetition count of the $j$-th workload component in the $i$-th time window; $M_h^{(i)}$ and $O_u^{(i)}$ are the aggregated values of the $h$-th performance metric and the $u$-th operator distribution in the $i$-th time window, respectively. Given component $C_j$, $m_{hj}'$ is the $h$-th performance metric, and $o_{uj}'$ is the $u$-th operator distribution; $T_j'$ is the duration of $C_j$; $n_u$, $n_h$, $v$ are the upper bounds of $u$, $h$, and $j$, respectively; $y$, $z$, $l$ are three constraint values.







\subsubsection{Programming Constraints}
To obtain a synthetic workload that is as realistic as possible, we impose the following constraints:
\begin{itemize}[leftmargin=*]

\item \textbf{Duration constraint.} To ensure that query execution completes within a time window and thus avoid affecting the subsequent workload, the total duration of all selected workload components, $x_j^{(i)}*T_j'$, must not exceed the time window length multiplied by the maximum concurrency of the system (usually set to the number of CPU cores). We represent this value as $l$.

\item  \textbf{Diversity constraint.} To prevent generating a workload by repetitions of a few workload components (which may occur when the features of these workload components are highly aligned with the targets), \sys\ needs to guide the workload synthesizer to combine different workload components for better diversity. This is achieved by limiting the number of repetitions of each workload component to a specified value $y$.

\item  \textbf{Total count constraint.} To avoid generating a workload consisting of massive short queries, \sys\ limits the total number of workload components used to $z$ (when the count of queries in the customer's workload is known, $z$ can be set accordingly).
\end{itemize}



\subsubsection{Integer Linear Programming}
\begin{figure*}
    \centering
    \includegraphics[width=\linewidth]{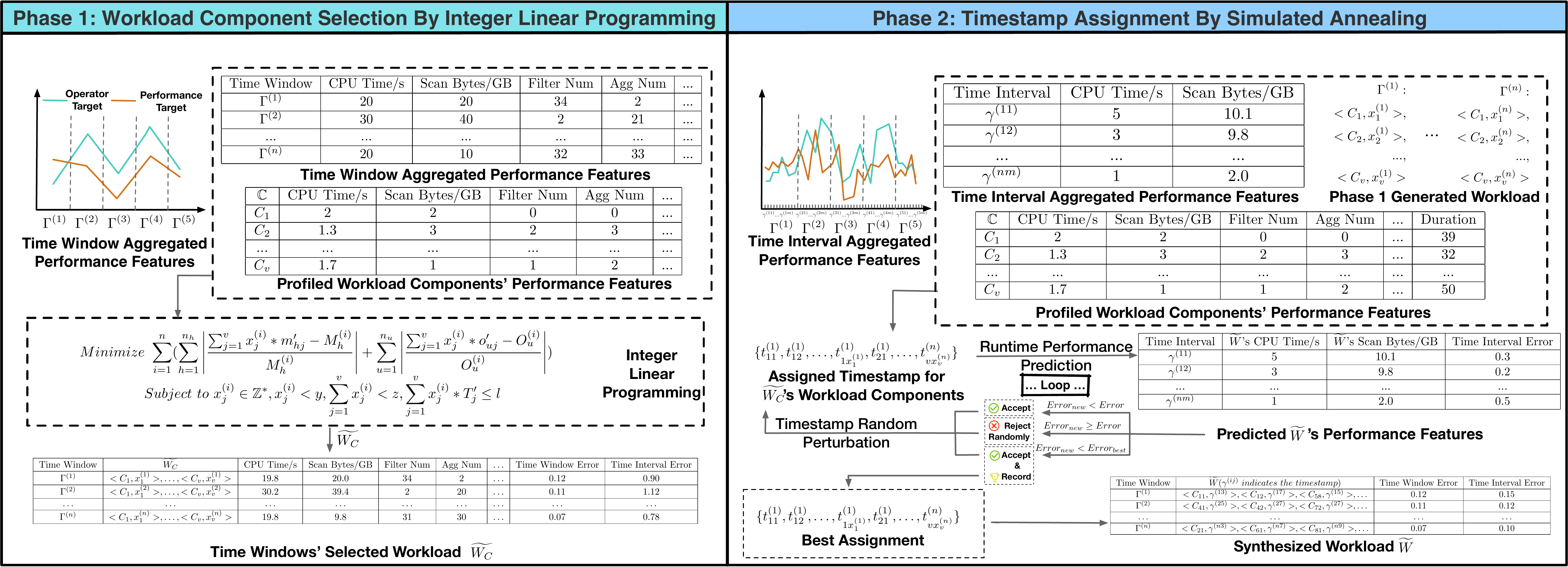}
    \vspace{-1.5em}
    \caption{Two Phase Workload Synthesis}
     \vspace{-1.5em}
    \label{fig:ILP}
\end{figure*}

Since we have the above optimization objectives, constraints, and decision variables, we can model the studied problem as an Integer Linear Programming problem. Subsequently, we can use any ILP Solver to find the optimal number of each workload component, thereby obtaining the synthesized workload with real statistics. As shown in Figure~\ref{fig:ILP}, given the target statistics (i.e., performance feature) for the \textit{n} time windows, the ILP solver aims to find a solution that approximates the performance feature in each time window. 

\begin{example}
Figure~\ref{fig:ILP} illustrates an example of our workload component selection method. For the time window $\Gamma^{(1)}$, both the target CPU Time and Scanned Bytes are 20, and the filter number and aggregate number are 34 and 2, respectively. By employing the ILP solver, we find a solution that repeats the workload component $C_1$ with the number of $x_{1}^{(1)}$, etc. This results in a workload with close performance features (i.e., 19.8 CPU Time and 20 Scanned Bytes, 34 filters, and 2 aggregates).
\end{example}







\subsection{Timestamp Assignment Phase}

After the first phase, \sys generates a workload that approximates the performance features within a coarse-grained time window. However, it still faces two key challenges

First, to achieve more fine-grained approximation with the synthetic workload, we should use smaller time windows whenever possible. However, applying smaller time windows will limit the space for combining workload components and will reduce the approximation accuracy of individual time windows. Moreover, longer queries may not be executed within a single time window, and thus cannot be applied to the linear synthesis process. This may result in the synthetic workload being dominated by short queries, compromising the effectiveness of the synthetic workload.
%
Second, the ILP-based solution focuses only on “\textit{which}” and “\textit{how many}” queries to send in each time window, neglecting “\textit{when}” they actually arrive. Sending queries at the very beginning or utilizing a Poisson distribution or a uniform distribution will all lead to differences compared to the distribution of actual query arrival time, thus limiting the effectiveness of synthetic workloads. 



To deal with the issues above, \sys\ proposes a fine-grained method to assign workload components' timestamps at the interval level, thereby better simulating the temporal dynamics of real workloads. Compared to approximating performance metrics like CPU Time and Scanned Bytes, approximating the operator distribution for each time interval is more challenging. Hence, we do not pursue approximating the operator distribution at the interval level.



\subsubsection{Concurrent Processing Assumption}
Normally, processing multiple queries in parallel affects the queries' execution time. Therefore, we need to effectively estimate the execution time of each query when multiple queries are executed simultaneously. Following ~\cite{autowlm}, we assume that an additional query linearly slows down existing queries. Namely, when a new query arrives with profiled duration $T$, the concurrency $P$ of the system increases by 1 ($P'=P+1$) and affects the new average duration by $T'=\frac{\frac{P'}{P}T+P(\hat{T}+\frac{1}{P})}{P'}$. The total delay caused by this average delay is allocated to each query according to the original query time distribution of all queries in the system, thus obtaining the estimated duration of all queries in the system at this time.

\subsubsection{Simulated Annealing}

To achieve a fine-grained generation, we model the timestamp assignment problem as an optimization problem, and the optimization goal is to minimize the total error of all time intervals in each time window $\Gamma^{(i)}$. Consequently, \sys\ develops a simulated annealing-based algorithm to assign the timestamp for each query (i.e., the query's starting time).  The algorithm performs in three steps. First, it initializes a timestamp assignment of workload components. Second, it randomly modifies the current assignment by adjusting one workload component's starting time to create a new solution. Third, it recursively checks if the new solution has a lower error. If this is the case, it accepts the current solution with a certain probability to avoid getting stuck in local optima. The probability of accepting a worse solution is controlled by a "temperature" parameter $V$ that gradually decreases as the algorithm progresses. The process repeats for a few iterations, gradually lowering the temperature and refining the solution, until a stopping criterion is met (e.g., no significant improvement over a set number $S$ of steps). 

\begin{example}
Figure~\ref{fig:ILP} illustrates an example of our timestamp assignment method. Particularly, $C_{11},C_{12},C_{51},...$ are the generated workload components for the time window $\Gamma^{(1)}$ from the first phase, in which $C_{11}$ and $C_{12}$ refer to the first selection and the second selection of $C_{1}$, respectively. To achieve fine-grained generation, \sys assigns the timestamp of the selected components by simulated annealing. For instance, it figures out the best timestamp for $C_{11}$ is $\gamma^{(13)}$ where $\gamma^{(13)}$ refers to the third time interval of $\Gamma^{(1)}$. As a result, we reduce the time interval error from 0.90 in the first phase to 0.15 in the second phase.
\end{example}

\subsection{Discussion: Query Level Matching}
Since \snowset\ and \redset\ are query-level statistics datasets, the most natural fitting approach is an actual query-to-query fitting. This involves identifying the workload component using a distance function tool or a combination of several workload components using ILP that is closest to the customer's query features and then using the combination as the synthesized workload. Such a method is equivalent to a special case where there is only one query within a "time window." This approach has notable advantages and disadvantages. The advantage is that it can directly send an alternative query at the exact moment the customer submits their query, eliminating the need for timestamp allocation and naturally achieving better fitting results for peaks. However, this method heavily relies on the similarity between target queries and workload components, leading to low accuracy and efficiency in practice.



\section{Workload Augmentation with LLMs}
\label{sec:LLM_gen}
The performance of the two-phase workload synthesis largely depends on the quality of the candidate set of workload components. 
Specifically, workload components from existing benchmarks may be insufficient to construct a synthetic workload that accurately reflects real traces due to the disparity between the original workload and the candidate workload. These disparities are often caused by different query patterns and underlying databases.
To address this, we propose to augment additional workload components of high quality  by leveraging LLMs.


\begin{figure}[t!]
    \centering
    \includegraphics[width=1\linewidth]{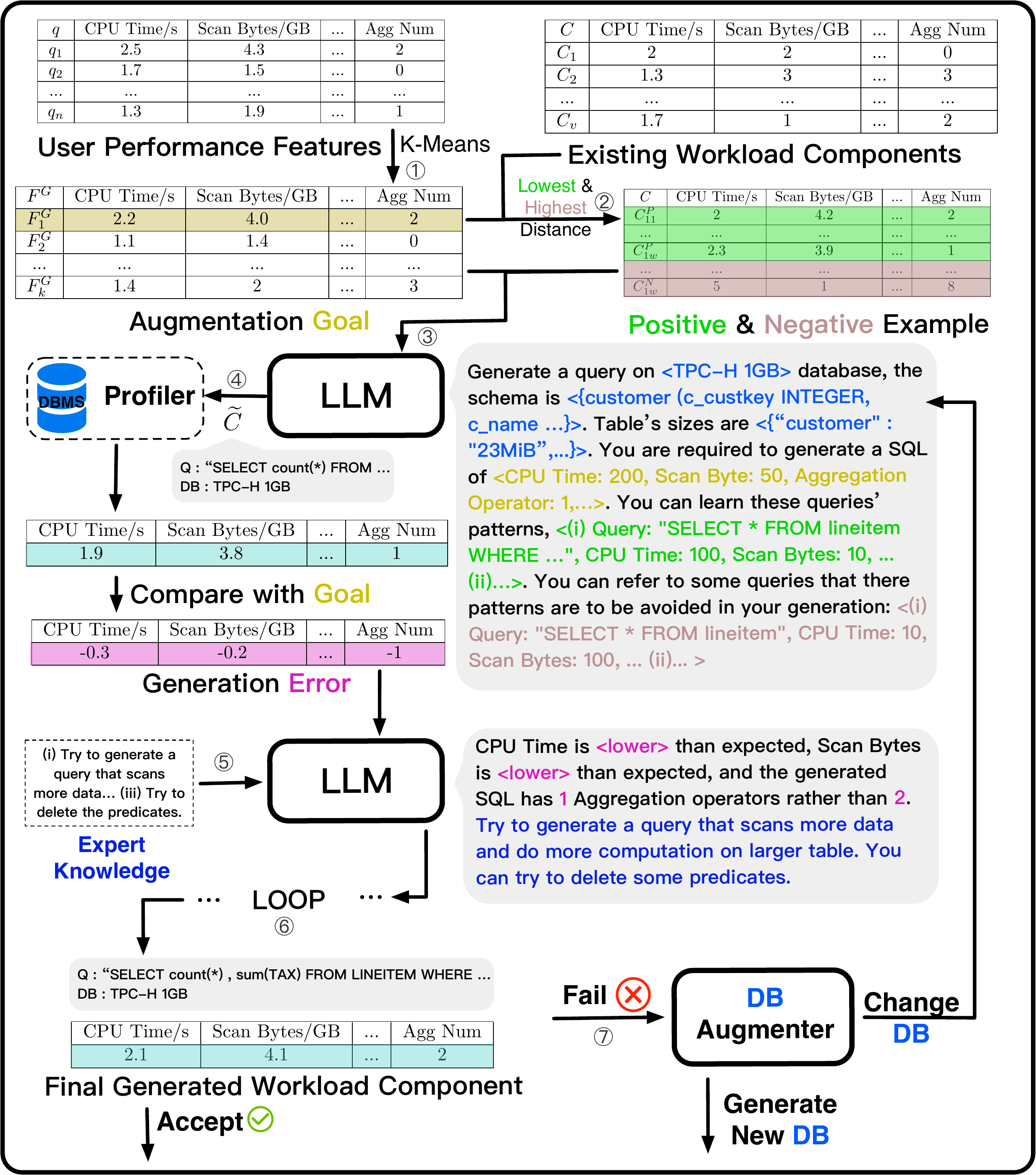}
    \vspace{-2em}
    \caption{LLM-enhanced Workload Augmentation}
        \vspace{-2em}
    \label{fig:LLM}
\end{figure}

When the workload $\widetilde{W}$ synthesized by the two-phase workload synthesis has rather low accuracy, it means that there is a discrepancy between the workload components and the target queries. If we can generate new queries on the benchmark database that produce performance features closer to the target, we can improve the performance of workload synthesis. \textit{The main challenge is to guide the LLM to understand the relationship between a query and its generation target~\cite{li2024llm}.} Our key idea is to prompt it with example hints. Specifically, \oursys\ operates in the following steps:

\textbf{Step 1: Target Generation.} Instead of generating new queries for all candidates, we only need to generate a few representative queries that can balance generation efficiency and approximation accuracy. \oursys\ leverages K-Means to cluster the query statistics $F_{j}$ within the time window and uses its centroids $F^G_j$ as the most "representative" generation targets. As shown in Figure \ref{fig:LLM}, the customer workload's statistics are clustered into $k$ centroids, and each centroid is set as a generation target.


\textbf{Step 2: Example retrieval.} Due to LLM's inability to directly understand the relationship between queries and their performance features in a zero-shot scenario, we need to provide LLM with informative cases for each generation target $F^G_j$ to assist it in generating new queries. We use the Euclidean distance to calculate the similarity between the generation target $F^G_j$ and the features of existing workload components $F_i'$. We take the N queries with the smallest and largest distances as positive and negative examples. We provide these cases to the LLM and ask it to learn the query patterns of the queries and avoid the query patterns of negative examples. For instance, for $F_1^G$, the positive and negative examples are marked in green and brown, respectively.

\textbf{Step 3: Prompt Generation.} We first pick a database that has the most positive examples corresponding to the generation target. Then we supply LLM with its metadata consisting of the schema and statistics, e.g., each table's size. We ask LLM to generate a query on the database and offer it with examples and their corresponding performance feature. 


\textbf{Step 4: Real-time Profiling. }After LLM generates a query, \sys\ profiles the generated queries and outputs the performance feature. The profiling process reveals the difference between the generation target and the generated query. 


\begin{table}[t!]
    \caption{Example Hints for Query and Database Augmenter}
      \vspace{-1em}
    \label{tab:hints}
\small
    \begin{tabular}{|>{\centering\arraybackslash}p{0.3\linewidth}|>{\centering\arraybackslash}p{0.6\linewidth}|} \hline 
        \textbf{Scenarios}& \textbf{Hints for LLM}\\ \hline 
        \textbf{Generated query has lower CPU Time \& lower Scanned Bytes}& (i) Try to generate a query that performs computation on a larger table. (ii) Try to delete some predicates to scan more data.\\ \hline 
        \textbf{Generated query has higher CPU Time \& lower Scanned Bytes}& (i) Scan more data while deleting some operators. (ii) Use more Inner Join operators to reduce the intermediate result size.\\ \hline 
        \textbf{Generated query has lower CPU Time \& higher Scanned Bytes}& (i) Use more Self Join operators to increase the size of intermediate results. (ii) Perform arithmetic operations on some columns. (iii) Scan a smaller table while adding more operators.\\ \hline 
        \textbf{CPU Time and Scanned Bytes are lower or higher} & Use or generate a benchmark database with a higher or lower Scale Factor.\\ \hline 
        \textbf{CPU Time/Scanned Bytes Ratio is lower or higher} & Use a benchmark database of higher or lower skewness, or select a database with a more complex schema. \\ \hline
    \end{tabular}
       \vspace{-2em}
\end{table}



\textbf{Step 5: Hint Construction.} We classify the generation cases into several scenarios, then we provide the LLM with the pre-defined hints. By providing these hints, we can obtain a prompt for regenerating a high-quality query. \autoref{tab:hints} illustrates several example hints. For instance, if the generated query has lower CPU Time and higher Scanned Bytes, \oursys\ guides LLM to add more calculations such as EXP or SUBSTRING on some columns, and use operators like \textit{Self Join} to increase the size of intermediate results.

\textbf{Step 6: Trial-and-Error Generation.} We guide the LLM to generate queries using a trial-and-error strategy. In each iteration, we revoke the profiling step and check the approximation error. If the error surpasses the acceptable threshold, we will regenerate the query in a few iterations.


\textbf{Step 7: Database Augmentation.} Due to the target's database being inaccessible, \oursys\ may not be able to generate the required queries on the existing benchmark database. In other words, it may not be possible to generate any query with performance feature $F_j^G$ on the current benchmark database. Therefore, we need to generate new databases when there are multiple failed attempts. We list the error scenarios in \autoref{tab:hints}. For example, when CPU Time and Scanned Bytes are consistently low or high, it indicates that the size of the selected database (Scale Factor) is too small or too large, and we need to adjust the Scale Factor. When there is no available database for the given Scale Factor, the database needs to be generated. Another example is when the ratio of CPU Time/Scanned Bytes is consistently too low compared to the target. This may suggest that the intermediate result size of the database may be insufficient. In such cases, we can improve the likelihood of creating larger intermediate results by using a database of higher skewness or selecting a database with more complex table structures and relationships (such as switching from TPC-H to TPC-DS).

\section{Experiments}
\label{sec:eval}






\subsection{Experiment Setup}

\begin{itemize}[leftmargin=*]
    \item \textbf{Platform and Implementation. }We use the same experiment setting as our baselines. For each server node, we use an 8C16G setting and we leverage Prometheus~\cite{Prometheus} to collect the metrics. For Redset's query-to-query baseline experiment, we run our evaluation on a 4-node ra3.4xlarge Redshift provisioned cluster and collect metrics via the statistic tables including stl\_wlm\_query, stl\_scan, and stl\_explain.  We use CBC Solver~\cite{cbc} as our ILP solver and Python SimAnneal Package~\cite{sa} as our SA solver. We use a state-of-the-art LLM model \footnote{We cannot disclose the model name due to Amazon policy.} for the Augmenter.

    \item \textbf{Workload Traces. }For \snowset, we cluster all the traces into five query arrival time patterns based on CAB~\cite{CAB}'s summarization. We randomly select a one-hour trace from each pattern's traces. These selected traces contain both short-term and long-term peak loads, ensuring good representativeness for the entire \snowset\ dataset. One of these traces is also used for ablation experiments. For \redset, we randomly select two one-hour traces for window-level experiments, and three 24-hour traces on four nodes for query-level experiments.



    \item \textbf{Workload Components.} We prepare TPC-H, TPC-DS, the Join Order Benchmark (JOB)~\cite{imdbbenchmark}, and YCSB~\cite{dey2014ycsb+} with various scale factors. JOB is a benchmark that contains 113 OLAP queries and a dataset collected from IMDB~\cite{imdb}. As for TPC-H, we use sizes of 500M, 1G, 2G, 5G, 9G. As for TPC-DS, we use sizes of 1G and 2G. As for Redset's 4-node experiments, we use sizes of 1G, 10G, 100G, 1T, and 3T for both TPC-H and TPC-DS. 
    
    
    \item \textbf{Evaluation Metrics.} For the window-level performance evaluation, we use Mean Absolute Error (MAE), Geometric Mean Absolute Percentage Error (GMAPE), and Geometric Mean Q-Error (GMQE) per time window as the evaluation metrics. We use the geometric mean of the relative error metrics instead of the arithmetic mean to avoid the influence of several extremely large relative error values. For window-level evaluation, $n$ denotes the number of time windows. For query-level evaluation, $n$ denotes the number of user queries.


    \begin{equation}
    \small
    \label{mae}
    \text{MAE= } \frac{1}{n} \sum_{i=1}^{n} \left | F^{(i)}-\widetilde{F^{(i)}}  \right |
    \end{equation}
    \begin{equation}
     \small
    \label{GMPE}
    \text{GMAPE= }  (\prod_{i=1}^{n}\left | \frac{F^{(i)}-\widetilde{F^{(i)}}}{F^{(i)}}   \right |+1)^{\frac{1}{n}}-1
    \end{equation}
    \begin{equation}
     \small
    \label{q-error}
        \text{GMQE= } 
        (\prod_{i=1}^{n}
        \max\left(\frac{F^{(i)}}{\widetilde{F^{(i)}}}, \frac{\widetilde{F^{(i)}}}{F^{(i)}}\right))^{\frac{1}{n} }
    \end{equation}

    \item \textbf{Baseline.} We also evaluate two baselines as follows: \\
    \textbf{(1) Stitcher~\cite{Stitcher}:} It synthesizes workloads by adjusting the sending concurrency and frequency of benchmark queries using Bayesian Optimization(BO)~\cite{frazier2018tutorial}. Since it is not open-sourced, we follow the instructions in ~\cite{Stitcher} and collect training data of various configurations of benchmark suite combinations based on the workload components we use. We train a linear model for each "workload type" (e.g. TPC-H 5G \& TPC-DS 1G), and use BO ~\cite{bayes-opt,bayes-opt-cons}) to find the solution. For each time window, we run 400 iterations of the BO processes to achieve the best accuracy. 
    
    
  
    \textbf{(2) CAB \cite{CAB}:} It creates a query pool and selects queries from the query pool to simulate the user workload's CPU Time. We use all our prepared workload components queries to construct the query pool for a fair comparison. Since it is open-sourced, we use their code to run the experiments.

    \item \textbf{Hyper parameter settings.} The time window $\Gamma$ is set to five minutes, and the time interval $\gamma$ is set to thirty seconds. For ILP, we set diversity constraint $y$ to 10, and the total count constraint $z$ is equal to twice the original user query count. For TA, we set the terminating criterion's no-improvement step number $S=100$ and we use a SA tool~\cite{sa} to automatically set the maximum and minimum temperatures. 
\end{itemize}





\subsection{Evaluation of \sys on Snowset}

\begin{table*}[t]
    \centering
    \caption{Evaluation of \sys on Snowset.}
     \vspace{-1em}
    \label{tab:main_comparison}
\resizebox{\textwidth}{!}{
\begin{tabular}{|c|ccc|ccc|cc|cc|cc|cc|}
\hline
\multirow{2}{*}{Methods} & \multicolumn{3}{c|}{CPU Time (s)}                                                           & \multicolumn{3}{c|}{Scanned Bytes (GB)}                                                & \multicolumn{2}{c|}{Filter Ratio}                  & \multicolumn{2}{c|}{Aggregate  Ratio}              & \multicolumn{2}{c|}{Join Ratio}                    & \multicolumn{2}{c|}{Sort Ratio}              \\ \cline{2-15} 
                         & \multicolumn{1}{c|}{MAE}            & \multicolumn{1}{c|}{GMAPE(\%)}         & GMQE       & \multicolumn{1}{c|}{MAE}           & \multicolumn{1}{c|}{GMAPE(\%)} & GMQE       & \multicolumn{1}{c|}{MAE}           & GMQE       & \multicolumn{1}{c|}{MAE}           & GMQE       & \multicolumn{1}{c|}{MAE}           & GMQE       & \multicolumn{1}{c|}{MAE}           & GMQE \\ \hline
Stitcher                 & \multicolumn{1}{c|}{64.89}         & \multicolumn{1}{c|}{110.16}         & 2.27& \multicolumn{1}{c|}{71.10}         & \multicolumn{1}{c|}{43.61} & 2.71& \multicolumn{1}{c|}{0.12}          & 1.35& \multicolumn{1}{c|}{0.12}          & 1.39& \multicolumn{1}{c|}{0.16}          & 1.02& \multicolumn{1}{c|}{0.02}          & 1.68\\ \hline
CAB                      & \multicolumn{1}{c|}{41.01}          & \multicolumn{1}{c|}{25.24}          & 1.25& \multicolumn{1}{c|}{24.83}         & \multicolumn{1}{c|}{45.43} & 1.86& \multicolumn{1}{c|}{0.73}          & 5.41& \multicolumn{1}{c|}{0.73}          & 5.25& \multicolumn{1}{c|}{0.66}          & 4.26& \multicolumn{1}{c|}{0.56}          & 5.05\\ \hline
\oursys   & \multicolumn{1}{c|}{\textbf{22.41}} & \multicolumn{1}{c|}{\textbf{17.37}} & \textbf{1.15}& \multicolumn{1}{c|}{\textbf{3.70}} & \multicolumn{1}{c|}{\textbf{12.43}}  & \textbf{1.25}& \multicolumn{1}{c|}{\textbf{0.00}} & \textbf{1.03} & \multicolumn{1}{c|}{\textbf{0.00}} & \textbf{1.03} & \multicolumn{1}{c|}{\textbf{0.00}} & \textbf{1.09} & \multicolumn{1}{c|}{\textbf{0.00}} & \textbf{1.00}\\ \hline
\end{tabular}%
}
\end{table*}





\stitle{Overall Evaluation.} 
Table~\ref{tab:main_comparison} reports the evaluation results of workload synthesis at the time window level over \snowset, reporting the average performance across five workload traces. For simplicity, we omit the GMAPE measurement of approximation errors for operator distributions.
The results indicate that \sys significantly outperforms the baselines, achieving up to $6\times$ lower GMAPE for performance metrics and much lower errors for operator distributions.
Specifically, \sys achieves a GMAPE of 17.37\% for CPU Time and 12.43\% for Scanned Bytes. The superior performance of \sys can be attributed to its workload synthesis approach, which effectively simultaneously approximate both performance metrics and operator distributions.

%

Specifically, since Stitcher can only adjust the sending frequency and concurrency of entire benchmark suites rather than individual queries, it lacks the flexibility needed to match real workload distributions precisely.
Moreover, CAB randomly selects queries to match the CPU Time of customer queries. However, this heuristic approach introduces substantial errors when dealing with skewed distributions, leading to worse CPU Time performance compared to \sys. Moreover, since CAB does not jointly optimize CPU Time and Scanned Bytes, it exhibits relatively high GMAPE error for Scanned Bytes when optimizing CPU Time.
%
%
The results further show that all baselines perform poorly in approximating operator distributions. In contrast, \sys consistently achieves near-zero MAE for all four operator types. This stems from two primary reasons: First, the baselines do not explicitly optimize for operator distributions, leading to substantial errors. Second, there exists a significant difference between the operator distributions in existing benchmarks and the real workload. Without augmenting new queries, the baselines struggle to bridge this gap.

In summary, the results demonstrate that, for real workload traces from \snowset, \sys achieves superior approximation of both performance metrics and operator distributions  and significantly outperforms existing methods.


\begin{figure*}[t]
    \centering
    \includegraphics[width=\textwidth]{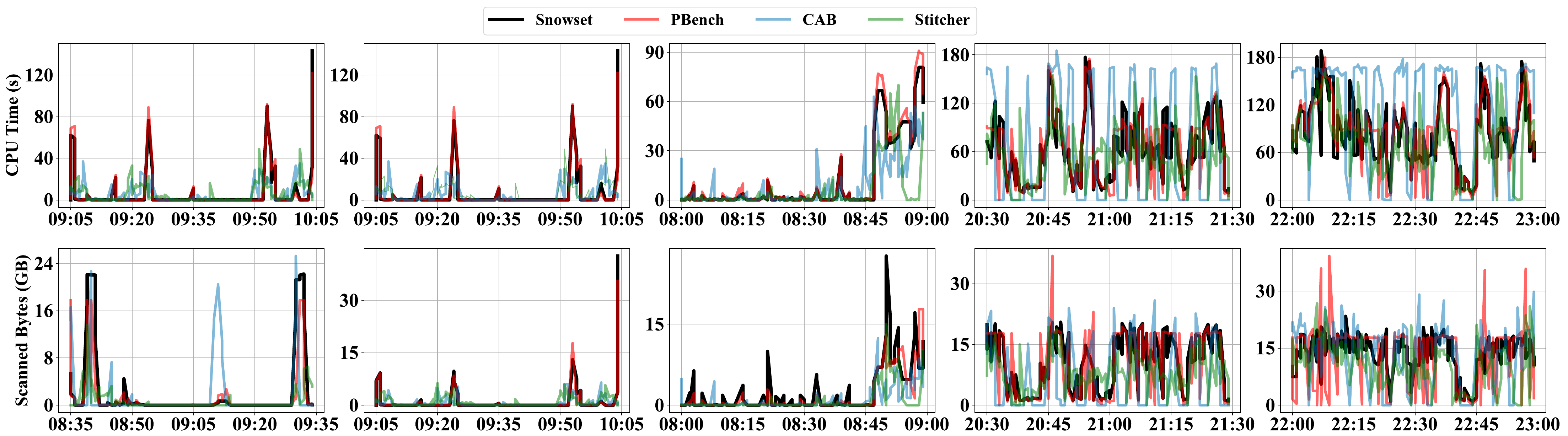}
    \vspace{-2em}
    \caption{Replayed Traces on Snowset with Different Methods }
    \vspace{-1em}
    \label{fig:snowset-all}
\end{figure*}

\begin{table}[t]
    \centering
    \caption{Results on Snowset (Time Interval)}
\label{tab:interval_snowset}
    \vspace{-1em}
\resizebox{\linewidth}{!}{%
\begin{tabular}{|c|ccc|ccc|}
\hline
\multirow{2}{*}{Methods} & \multicolumn{3}{c|}{CPU Time}                                           & \multicolumn{3}{c|}{Scanned Bytes}                                          \\ \cline{2-7} 
                         & \multicolumn{1}{c|}{MAE (s)} & \multicolumn{1}{c|}{GMAPE (\%)} & GMQE & \multicolumn{1}{c|}{MAE (GB)} & \multicolumn{1}{c|}{GMAPE (\%)} & GMQE \\ \hline
Stitcher                 & \multicolumn{1}{c|}{20.29}        & \multicolumn{1}{c|}{163.10}          &         4.41& \multicolumn{1}{c|}{3.37}         & \multicolumn{1}{c|}{82.85}          &         2.71\\ \hline
CAB                      & \multicolumn{1}{c|}{32.13}   & \multicolumn{1}{c|}{154.18}   & 10.57& \multicolumn{1}{c|}{4.52}     & \multicolumn{1}{c|}{80.57}    & 4.73\\ \hline
PBench                   & \multicolumn{1}{c|}{\textbf{7.75}}    & \multicolumn{1}{c|}{\textbf{36.05}}     & \textbf{1.49}& \multicolumn{1}{c|}{\textbf{2.32}}     & \multicolumn{1}{c|}{\textbf{25.23}}     & \textbf{1.56}\\ \hline
\end{tabular}%
}
    \vspace{-2em}
\end{table}

\stitle{Time Interval Results. }
We also evaluate the performance of the approaches on time interval results. As shown in \autoref{tab:interval_snowset}, \sys achieves more than $3\times$ lower GMAPE compared to CAB and Stitcher for CPU Time and Scanned Bytes. Due to the small sample size within each time interval, errors can be influenced by singular values, leading to a higher mean error. Nevertheless, \sys effectively captures peak patterns.
Figure~\ref{fig:snowset-all} compares the original \snowset traces with the replayed performance traces of the synthetic workloads. 
\sys consistently achieves better approximation performance across different workload patterns, including high spikes, low peaks, and sustained peaks. In contrast, CAB and Stitcher struggle to approximate certain peaks and exhibit larger errors due to their coarse granularity and limited adaptability.


It is worth noting that \sys may occasionally exhibit slight deviations in peak positions. This is primarily due to inaccuracies in estimating execution times under concurrent query processing, which is a challenge that affects all methods, including the baselines. While we apply a linear slow-down model to approximate these variations, completely eliminating such mismatches is infeasible. Fortunately, as these deviations result only in minor peak shifts or slight peak flattening within a few seconds, it has an insignificant impact on the overall performance.

\stitle{Efficiency Evaluation.} 
We also evaluate the generation time of each method to compare their efficiency. CAB synthesizes workloads within milliseconds, as it simply selects queries randomly without requiring training or prediction. Stitcher incurs minimal time for model training, as it only fits a few linear models, but requires an average of 122 minutes to assemble the workload. In contrast, \sys significantly improves efficiency, taking an average of 6 minutes for ILP and 8 minutes for TA, resulting in a total synthesis time of 14 minutes—making it $8\times$ faster than Stitcher.
It is worth noting that for real-time workload synthesis and replay, the synthesis time must remain within the length of the workload trace (60 minutes in this case) to ensure uninterrupted processing. Therfore, \sys meets this requirement, demonstrating its practicality for online workload synthesis and benchmarking.

%

\subsection{Ablation Study of \sys}

 \begin{figure}[t!]
      \label{fig:ablation}
     \begin{subfigure}[t]{\linewidth}
         \centering
    \includegraphics[width=\linewidth]{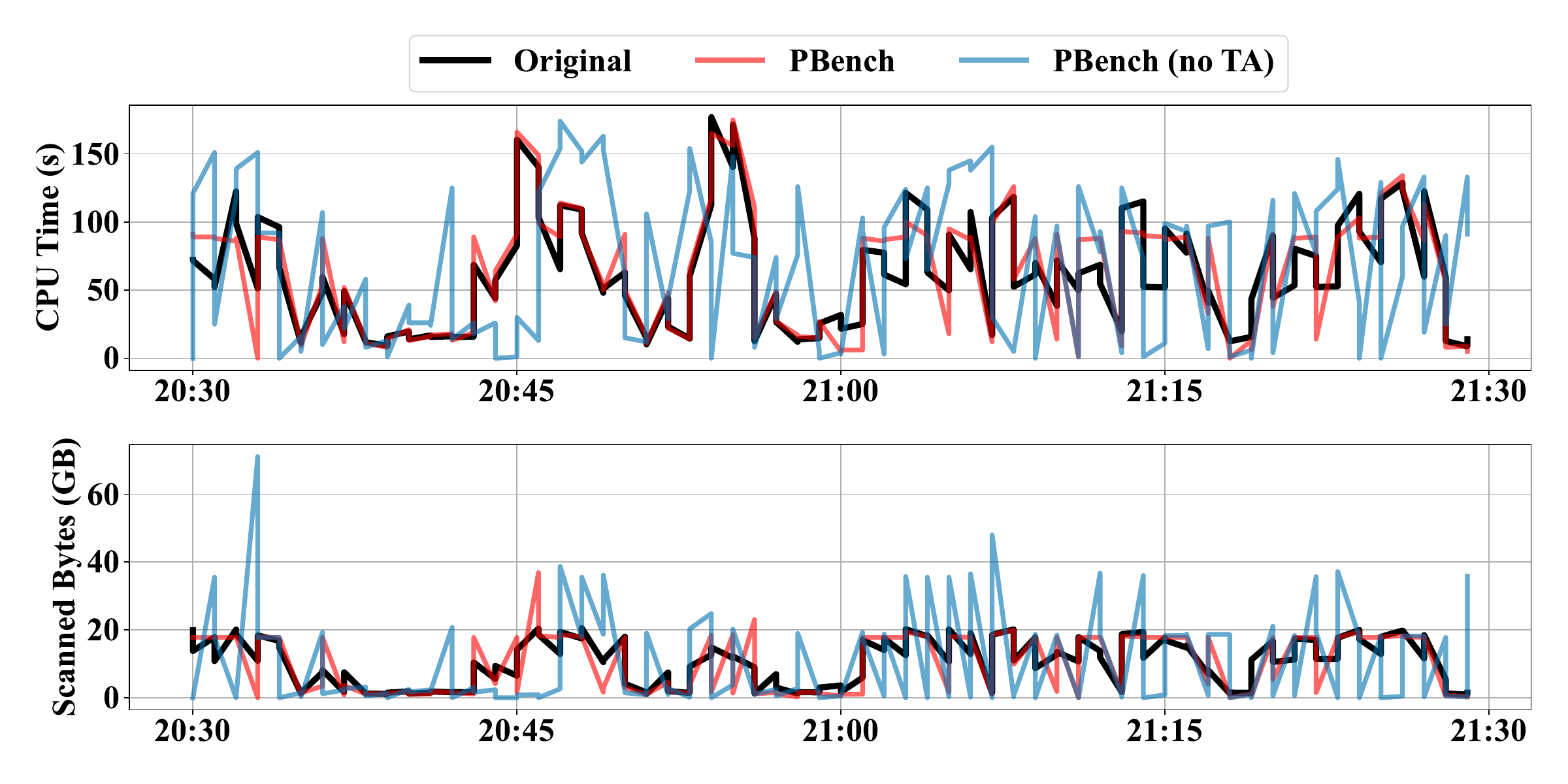}
    \vspace{-2em}
    \caption{Replayed Traces of  w/o Timestamp Assignment on Snowset}

    \label{fig:ablation-sa}
     \end{subfigure}
     \begin{subfigure}[t]{ \linewidth}
         \centering
    \includegraphics[width=\linewidth]{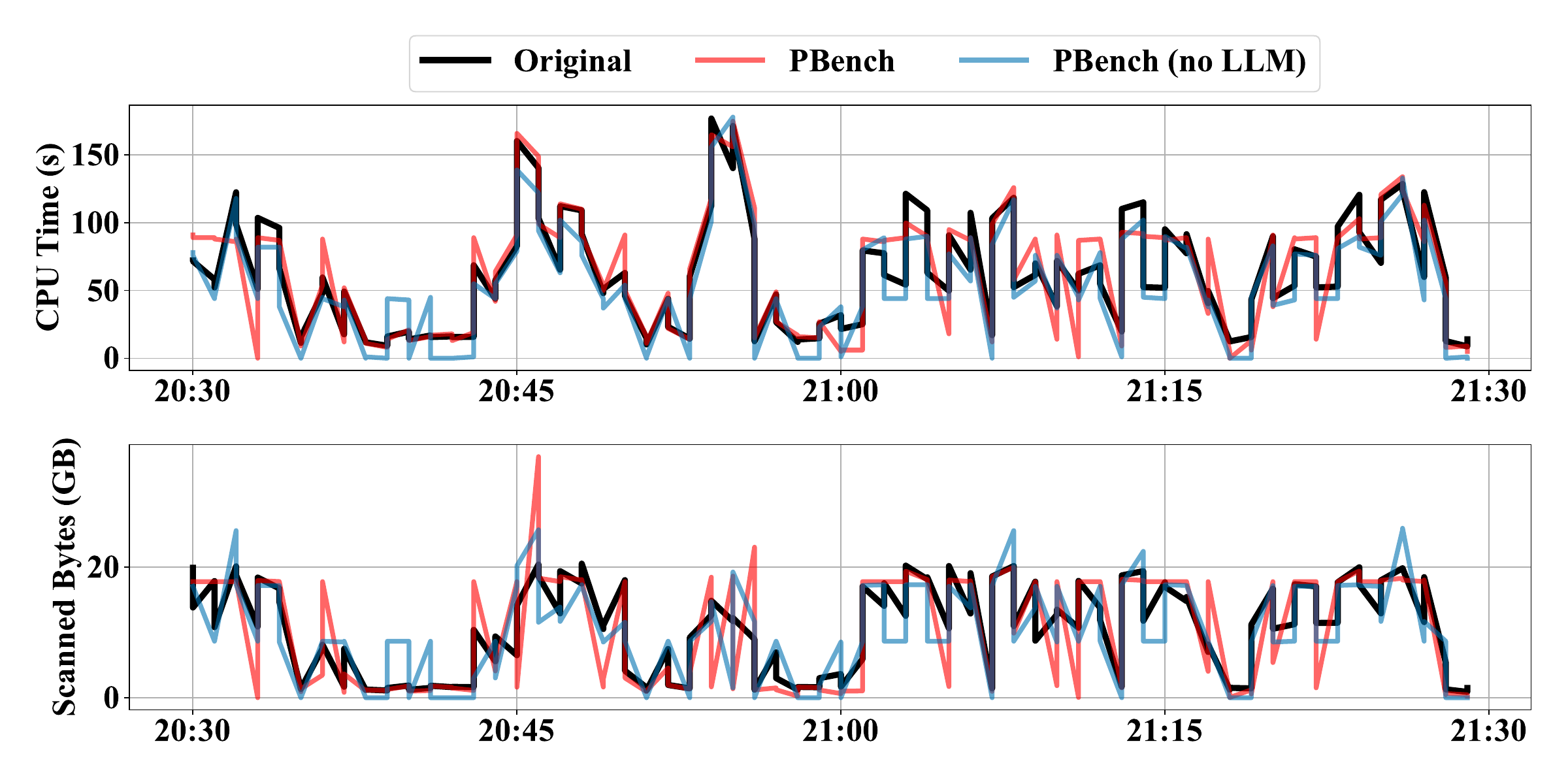}
    \vspace{-2em}
    \caption{Replayed Traces of  w/o LLM Augmenter on Snowset} 
    \label{fig:ablation-llm}
     \end{subfigure}
         \vspace{-1em}
       \caption{Ablation Study Replayed Traces on Snowset}
           \vspace{-1em}
 \end{figure}

\begin{table*}[t!]
    \centering
    \caption{Ablation Study Results on Snowset}
    \vspace{-1em}
    \label{tab:woLLMop}
\resizebox{\textwidth}{!}{
\begin{tabular}{|c|ccc|ccc|cc|cc|cc|cc|}
\hline
\multirow{2}{*}{Methods} & \multicolumn{3}{c|}{CPU Time (s)}                                                           & \multicolumn{3}{c|}{Scanned Bytes (GB)}                                                & \multicolumn{2}{c|}{Filter Ratio}                  & \multicolumn{2}{c|}{Aggregate  Ratio}              & \multicolumn{2}{c|}{Join Ratio}                    & \multicolumn{2}{c|}{Sort Ratio}              \\ \cline{2-15} 
                         & \multicolumn{1}{c|}{MAE}            & \multicolumn{1}{c|}{GMAPE(\%)}         & GMQE       & \multicolumn{1}{c|}{MAE}           & \multicolumn{1}{c|}{GMAPE(\%)} & GMQE       & \multicolumn{1}{c|}{MAE}           & GMQE       & \multicolumn{1}{c|}{MAE}           & GMQE       & \multicolumn{1}{c|}{MAE}           & GMQE       & \multicolumn{1}{c|}{MAE}           & GMQE \\ \hline
\sys\ (no LLM)& \multicolumn{1}{c|}{75.05}          & \multicolumn{1}{c|}{12.16}          & 1.14& \multicolumn{1}{c|}{7.33}         & \multicolumn{1}{c|}{7.98} & 1.08& \multicolumn{1}{c|}{0.93}          & 9.84& \multicolumn{1}{c|}{0.98}          & 9.99& \multicolumn{1}{c|}{0.27}          & 2.55& \multicolumn{1}{c|}{0.64}          & 6.28\\ \hline
\sys\ (no TA)& \multicolumn{1}{c|}{65.88} & \multicolumn{1}{c|}{10.48} & 1.10& \multicolumn{1}{c|}{0.52} & \multicolumn{1}{c|}{0.49}  & 1.00& \multicolumn{1}{c|}{0.01} & 1.02& \multicolumn{1}{c|}{0.00} & 1.00& \multicolumn{1}{c|}{0.02} & 1.14& \multicolumn{1}{c|}{0.01} & 1.00\\ \hline
\oursys   & \multicolumn{1}{c|}{\textbf{23.72}} & \multicolumn{1}{c|}{\textbf{4.40}} & \textbf{1.04}& \multicolumn{1}{c|}{\textbf{0.44}} & \multicolumn{1}{c|}{\textbf{0.31}}  & \textbf{1.00}& \multicolumn{1}{c|}{\textbf{0.01}} & \textbf{1.02}& \multicolumn{1}{c|}{\textbf{0.00}} & \textbf{1.00}& \multicolumn{1}{c|}{\textbf{0.02}} & \textbf{1.14}& \multicolumn{1}{c|}{\textbf{0.01}} & \textbf{1.00}\\ \hline
\end{tabular}%
    \vspace{-3em}
}
\end{table*}

In this section, we present an ablation study of \oursys, focusing on two key innovations: the fine-grained Timestamp Assignment (TA) phase and the LLM-based Component Augmentation.



\textit{(1) w/o Timestamp Assignment.} \autoref{fig:ablation-sa} shows the results of \sys's w/o Timestamp Assignment (TA) results. Since our baselines have no timestamp assignment step, we randomly send workload components within the time window. It can be seen that when we do not assign timestamps (i.e., performing the first phase only), the fitting result shows a significant mismatch. With timestamp assignment, the results for performance metrics are significantly improved (i.e., $2.4\times$ and $1.6\times$ better GMAPE on CPU Time and Scanned Bytes, respectively). Note that TA has no impact on the operator results as the ratio is fixed in a time window regardless of the queries' order or the length of the time window.




\textit{(2) w/o Component Augmentation.} \autoref{fig:ablation-llm} shows the comparison results with and without the LLM-based workload components augmenter. It can be observed that the workload component added by LLM significantly improves the fitting accuracy. Since there are significant differences between benchmark queries and target queries, it is imperative to generate new candidate queries. The results in \autoref{tab:woLLMop} indicate that the added workload components not only improve the accuracy of performance metrics fitting but also significantly improve the accuracy of operator ratio fitting. The traditional benchmark queries typically have many computational operators, differing from the distribution of customers' operators. By adding workload components through our LLM-enhanced augmenter, the fitting error can be significantly reduced.


\subsection{Evaluation of \oursys\ on Redset}

\begin{table*}[t!]
    \centering
            \vspace{-1em}

    \caption{Results on Redset (Window Level)}
        \vspace{-1em}
    \label{tab:redset_comparison}
\begin{tabular}{|c|c|c|c|c|c|c|c|c|c|c|c|c|}
\hline
\multirow{2}{*}{Methods} & \multicolumn{3}{c|}{CPU Time (s)}                                                           & \multicolumn{3}{c|}{Scanned Bytes (GB)}                                                                   & \multicolumn{3}{c|}{Join Num} & \multicolumn{3}{c|}{Aggregate Num} \\   \cline{2-13}&MAE            & GMAPE (\%)        & GMQE       & MAE      &GMAPE (\%) & GMQE                             & MAE& GMAPE &GMQE        & MAE& GMAPE&GMQE                   \\ \hline
Stitcher                 & \multicolumn{1}{c|}{18.17}         & \multicolumn{1}{c|}{27.09}         & 1.50& \multicolumn{1}{c|}{8.50}         & \multicolumn{1}{c|}{\textbf{61.65}} &           3.42& 56.45& 337.82& 5.28& 135.21& 74.56&          5.81\\ \hline
CAB                      & \multicolumn{1}{c|}{8.29}          & \multicolumn{1}{c|}{25.70}          & 1.26& \multicolumn{1}{c|}{10.89}         & \multicolumn{1}{c|}{138.75} &           4.17& 187.45& 95.52& 36.29& 55.48& 188.43&10.72\\ \hline
\oursys   & \multicolumn{1}{c|}{\textbf{7.52}} & \multicolumn{1}{c|}{\textbf{24.49}} & \textbf{1.24}& \multicolumn{1}{c|}{\textbf{6.83}} & \multicolumn{1}{c|}{97.16} & \textbf{2.53} & \textbf{0.59}& \textbf{6.56}&  \textbf{1.06} & \textbf{0.47}& \textbf{0.46}&\textbf{1.00} \\ \hline
\end{tabular}%

\end{table*}

 \begin{figure*}[t!]
    \centering
    \includegraphics[width=\textwidth]{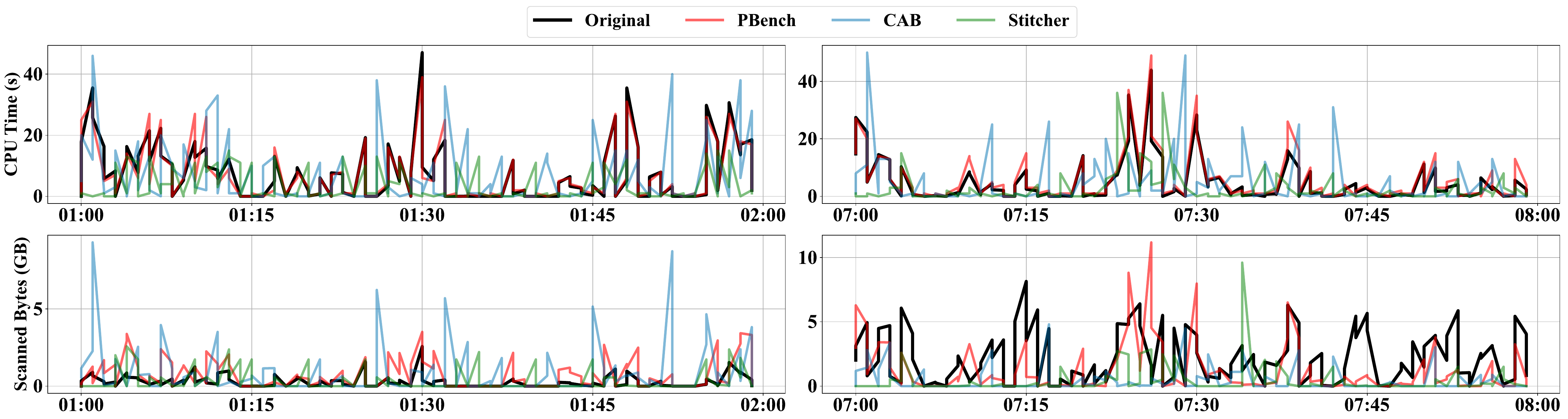}
    \vspace{-2em}
    \caption{Replayed Traces on Redset}
    
    \label{fig:redset-all}
    \vspace{-1em}
\end{figure*}

\stitle{Window Level Result.}\label{window_redset} \autoref{fig:redset-all}  and \autoref{tab:redset_comparison} show the results of \oursys\ and baselines on three Redset traces.  \autoref{tab:redset_comparison}
 shows that \oursys\ outperforms baselines as well. For CPU Time, \sys\ achieves better GMAPE than Stitcher and CAB. As for Scanned Bytes, \sys\ outperforms Stitcher and CAB in terms of MAE. As for operator numbers like Aggregation and Join number, \sys\ excels due to its multi-objective optimization and workload component augmenter, thus it outperforms all the baselines and achieves a GMQE of 1.00 for the aggregation number. \autoref{fig:redset-all} shows that \sys\ can fit almost every peak of the customer trace, indicating that the timestamp assignment method works well on Redset.  We also evaluate the performance metrics matching accuracy in the interval level. As shown in \autoref{tab:interval_redset}, we achieve more than $2.5\times$ higher accuracy in GMAPE than the baselines for CPU Time.


 
 

\begin{table*}[!t]
    \centering
    \caption{Results on Redset (Query Level)}
     \vspace{-1em}
     \resizebox{0.95\linewidth}{!}{%
    \begin{tabular}{|c|c|c|c|c|c|c|c|c|c|c|c|c|c|}
    \hline
        \multirow{2}*{Trace}& \multirow{2}*{Methods}& \multicolumn{3}{c|}{CPU Time (s)}& \multicolumn{3}{c|}{Scanned Bytes (GB)}& \multicolumn{3}{c|}{Join Num}&\multicolumn{3}{c|}{Aggregation Num}\\
        \cline{3-14}& & MAE & GMAPE& GMQE& MAE & GMAPE& GMQE& MAE & GMAPE& GMQE& MAE & GMAPE&GMQE\\ \hline
        \multirow{3}*{T1}& one-to-one& 26.61& 32.98&   2.72
& 8.86
& 41.54&   1.45
& 0.72& 55.43& 1.02
& 0.47& 10.49& 1.11
\\ 
        \cline{2-14}& one-to-many& \cellcolor{gray!40}{26.54}& \cellcolor{gray!40}{32.27}& \cellcolor{gray!40}{2.70}& \cellcolor{gray!40}{8.32}
& \cellcolor{gray!40}{40.92}&  \cellcolor{gray!40}{1.44}& \cellcolor{gray!40}{0.69}& \cellcolor{gray!40}{54.93}& \cellcolor{gray!40}{1.02}& \cellcolor{gray!40}{0.38}& \cellcolor{gray!40}{9.62}& \cellcolor{gray!40}{1.09}\\ \hline
        \multirow{3}*{T2}& one-to-one
& 15.3& 25.91& 1.56
& 0.03
& 109.27& 2.07
& 0.49& 23.61& 1.05
& 0.34& 12.84& 1.09
\\
        \cline{2-14}& one-to-many
& \cellcolor{gray!40}{13.36}& \cellcolor{gray!40}{25.00}& \cellcolor{gray!40}{1.54}& \cellcolor{gray!40}{0.02}
& \cellcolor{gray!40}{106.90}& \cellcolor{gray!40}{2.04}& \cellcolor{gray!40}{0.45}& \cellcolor{gray!40}{22.53}& \cellcolor{gray!40}{1.05}& \cellcolor{gray!40}{0.25}& \cellcolor{gray!40}{11.00}& \cellcolor{gray!40}{1.07}\\ \hline
        \multirow{3}*{T3}& one-to-one
& 223.64& 35.75& 1.89
& 0.02
& 17.85& 1.88
& 0.72& 65.36& 1.00& 0.07& 6.97&1.02
\\
        \cline{2-14}& one-to-many
& \cellcolor{gray!40}{191.23}& \cellcolor{gray!40}{35.65}& \cellcolor{gray!40}{1.88}& \cellcolor{gray!40}{0.02}
& \cellcolor{gray!40}{17.79}& \cellcolor{gray!40}{1.18}& \cellcolor{gray!40}{0.72}& \cellcolor{gray!40}{65.36}& \cellcolor{gray!40}{1.00}& \cellcolor{gray!40}{0.07}& \cellcolor{gray!40}{6.94}& \cellcolor{gray!40}{1.02}\\ \hline
    \end{tabular}
    }
  \vspace{-.5em}
    \label{tab:redshift}
\end{table*}

\begin{table}[t!]
    \centering
        \vspace{-0.5em} 

    \caption{Results on Redset (Time Interval)}
    \vspace{-1em} 
\resizebox{\linewidth}{!}{%
\begin{tabular}{|c|ccc|ccc|}
\hline
\multirow{2}{*}{Methods} & \multicolumn{3}{c|}{CPU Time (s)}                                           & \multicolumn{3}{c|}{Scanned Bytes (GB)}                                          \\ \cline{2-7} 
                         & \multicolumn{1}{c|}{MAE} & \multicolumn{1}{c|}{GMAPE (\%)} & GMQE & \multicolumn{1}{c|}{MAE} & \multicolumn{1}{c|}{GMAPE (\%)} & GMQE \\ \hline
Stitcher                 & \multicolumn{1}{c|}{6.03}        & \multicolumn{1}{c|}{197.68}          &         8.24& \multicolumn{1}{c|}{1.21}         & \multicolumn{1}{c|}{90.94}          &         4.09\\ \hline
CAB                      & \multicolumn{1}{c|}{7.41}   & \multicolumn{1}{c|}{230.68}   & 7.41& \multicolumn{1}{c|}{1.39}     & \multicolumn{1}{c|}{128.87}    & 4.67\\ \hline
PBench                   & \multicolumn{1}{c|}{\textbf{1.43}}    & \multicolumn{1}{c|}{\textbf{77.07}}     & \textbf{1.82}& \multicolumn{1}{c|}{\textbf{1.06}}     & \multicolumn{1}{c|}{\textbf{85.87}}     & \textbf{2.64}\\ \hline
\end{tabular}%
}
\label{tab:interval_redset}
    \vspace{-1em}
\end{table}

\stitle{Query Level Results.} We conduct a query-level synthesizing experiment on Redset's 3 provisioned traces (ids 73, 132, and 181 in~\cite{redset}), all for 4-node clusters. We choose queries received during 24 hours on 03/01/2024. The workload volume shows significant variation across these traces: 1,694 queries for trace 73; 648,649 for trace 132; and 76,789 for trace 181. We generate 1000 queries on the TPC-H database as candidate workload components. For a query's performance feature $F_j$ in customer traces, we employ the ILP to synthesize a workload to replace $q_j$ as the one-to-many setting. For the one-to-one setting, we limit the query counts of the ILP to one and degenerate into finding the workload component with the closest Euclidean distance to $ F_j$. Since we can send the workload components at the exact time of $q_j$, we do not need to perform the timestamp assignment in this experiment. As shown in \autoref{tab:redshift}, we found that the one-to-many approach with ILP is better than the one-to-one approach in most cases, indicating that the flexibility provided by workload combining is crucial. Also, when we only use benchmark queries without adding new queries, the experimental results deteriorate significantly, by up to $10\times$. 



\begin{figure}
      \label{fig:redset}
     \begin{subfigure}[t]{\linewidth}
         \centering
         \includegraphics[page=1, width=0.8\linewidth]{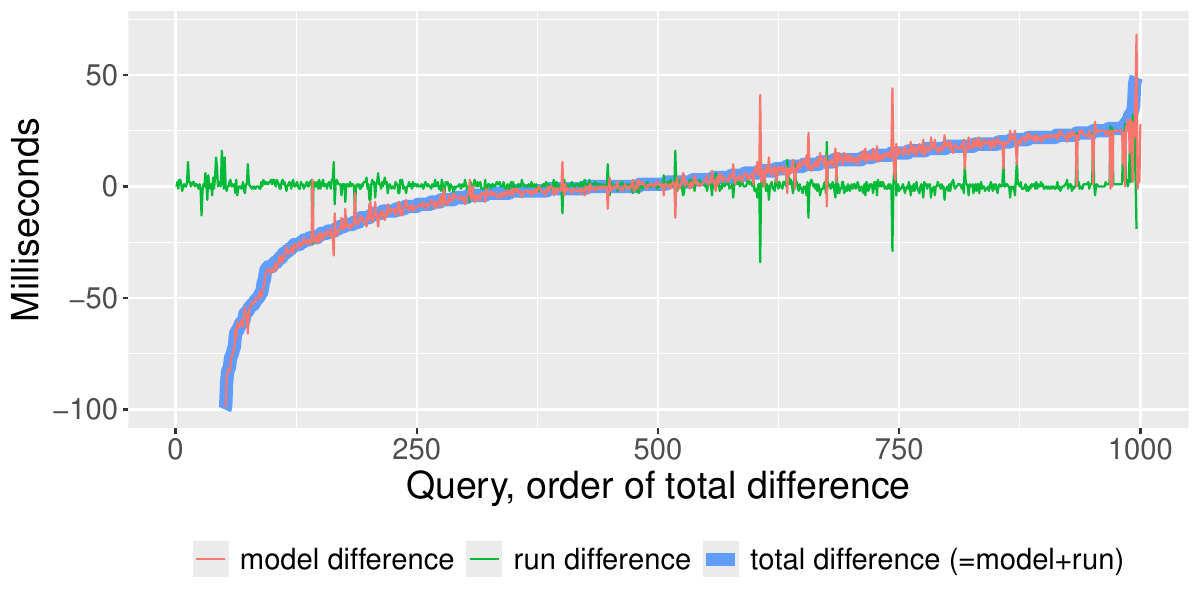}
         \vspace{-1.1em}
         \caption{Mismatch differences for execution time.}
         \label{fig:difflines_execTime}
     \end{subfigure}
     \begin{subfigure}[t]{\linewidth}
         \centering
         \includegraphics[page=2, width=0.8\linewidth]{figures/diff_lines_132.pdf}
                  \vspace{-1.1em}

         \caption{Mismatch differences for Scanned Bytes.}
         \vspace{-2.5em}
         \label{fig:difflines_scannedBytes}
     \end{subfigure}
       \vspace{-1em}
       \caption{Error Evaluation Results on Redset (Query Level)}
       \vspace{-1.5em}
 \end{figure}

\stitle{Query Level Error Evaluation.} Figure~\ref{fig:difflines_scannedBytes} depicts the end-to-end error evaluation for execution time and Scanned Bytes in the query-level experiment. The thick blue total difference line shows the end-to-end difference between \sys\ and \redset. The total difference depends on two steps: (1) the modeling step, i.e., solving the optimization problem given the pool of queries, and (2) the run/replay step, where we actually execute the queries. These two steps are represented by the red line and the green line. When added together, they result in the blue total difference. The x-axis represents the Redset queries in increasing order of total difference, which explains its monotonicity. We found that except for a very small number of queries, the run error is consistently low, and the model error dominates in most cases. This suggests using linear models for prediction is practical in real-world scenarios. For most cases, the total error is low, indicating that combining workload components by ILP performs well in the query-to-query method.

\section{Related Work}
\label{sec:related}

\noindent \textbf{Benchmark-Based Workload Synthesis.} There have emerged several works aiming to synthesize new workloads based on existing benchmarks. Particularly, Wan et al. developed Stitcher~\cite{Stitcher} to combine entire database benchmarks such as TPC-H's 22 queries and YCSB~\cite{YCSB}'s workload. The system comprises a predictor, a generator, and an integrator, with linear regression models trained for each benchmark combination to predict performance metrics; the generator employs Bayesian optimization to minimize the mean squared percentage error against the original workload's performance metrics; the integrator extracts representative time periods from the original workload and uses these to guide the generator in searching for benchmark settings. Renen et al.~\cite{CAB} introduced the CAB tool for synthesizing analytical workloads, which sets database scales based on log-normal distributions. CAB allocates CPU time budgets proportional to the actual sizes of the databases. It schedules query arrivals by segmenting the benchmark execution into time periods, where CPU time consumption is proportional to the number of query requests. CAB summarizes five typical query arrival patterns corresponding to different application scenarios and simulates a query flow with peaks and idle times by randomly selecting queries from a pool. It estimates their CPU time requirements while adhering to Poisson-distributed time intervals between queries. MTD-DS~\cite{MTDDS} extends the original benchmark to adapt to more complex query, SLA sensitive, and multi-tenant. Unfortunately, these works cannot achieve high accuracy as they employ a coarse-grained fitting strategy and neglect the operator ratio, while \oursys\ develops a fine-grained method to consider both performance metrics and operator distributions.

\noindent \textbf{Constraint-Aware Query Generation.} This line of work aims to create queries that conform to cardinality constraints based on given database tables. For instance, Bruno's climbing algorithm addresses database query cardinality constraints by optimizing predicate parameters \cite{bruno}, assuming independence, and using logarithmic relationships to form linear equations, which are solved using recursive least squares estimation. The algorithm employs hill-climbing with various parameter adjustment strategies and boundary clamping to refine searches and avoid local minima. Mishra's TQGen algorithm \cite{TQGen} reduces the search space for query parameters by determining parameter bounds and prioritizing exploration based on constraint satisfaction and uniformity, using a scoring system that considers both the number of constraints and the standard deviation of parameter values. Zhang \cite{LearnedSQLGen} et al. developed the cardinality-aware query generation framework, which treats the query generation problem as a sequential decision process and employs reinforcement learning (RL) techniques for solving it. Mueller et al. present novel approaches to represent queries and select predicates in a feature vector \cite{muller2023enhanced}. However, DL-based or RL-based approaches are time-consuming and tedious to train~\cite{li2021machine}. Moreover, they do not consider performance metrics.

\section{Conclusion}
\label{sec:con}
In this paper, we have introduced the problem of \emph{workload synthesis with real statistics} and proposed \sys, a novel workload synthesizer that constructs synthetic workloads by strategically selecting and combining workload components from existing benchmarks.
To achieve this, \sys incorporates effective techniques for component selection and timestamp assignment, and leverages an LLM-enhanced approach for component augmentation.
Extensive experiments on real cloud workload traces demonstrate that \sys significantly reduces approximation errors compared to state-of-the-art methods, effectively bridging the gap between synthetic workloads and real-world cloud workloads.

%

\clearpage

\balance

\bibliographystyle{ACM-Reference-Format}
\bibliography{references}

\end{document}